\documentclass[]{article}
\usepackage[T1]{fontenc}% optional T1 font encoding

\usepackage[caption=false,font=footnotesize]{subfig}
\usepackage[export]{adjustbox}

\usepackage{amsmath}
\usepackage{amssymb}
\usepackage{latexsym,cite}
\usepackage{graphicx}
\usepackage{svg}
\usepackage{epstopdf}
\usepackage{multirow}

\usepackage[cmintegrals]{newtxmath}

\usepackage{balance}

\usepackage{graphicx}
\usepackage{epstopdf}
\usepackage{cite}
\usepackage{multirow}
\usepackage{cleveref}
\usepackage{balance}
\usepackage{footnote}
\makesavenoteenv{tabular}
\usepackage{array}
\usepackage{dblfloatfix}
\usepackage{color}

\usepackage{graphicx}
\usepackage{epstopdf}
\usepackage{cite}
\usepackage{multirow}
\usepackage{cleveref}
\usepackage{balance}
\usepackage{footnote}
\makesavenoteenv{tabular}
\usepackage{array}
\usepackage{dblfloatfix}
\usepackage{color}

\usepackage{blindtext}
\date{}
\usepackage[left=2.30cm, right=2.30cm, top=2.30cm, bottom=2.30cm]{geometry}

\title{Millimeter-wave  Multi-mode Circular Antenna Array\\for Uni-cast Multi-cast and OAM Communication}

\author{
	Stylianos~D.~Assimonis, M.~Ali~Babar~Abbasi, and Vincent~Fusco
	\thanks{The authors are with the Institute of Electronics, Communications and Information Technology (ECIT), Queen's University Belfast, Belfast, BT3 9DT, U.K., (e-mail: \{s.assimonis,~m.abbasi,~v.fusco\}@qub.ac.uk).}
}

\begin{document}
	
	\maketitle
	
\begin{abstract}
	This paper investigates multi-mode and orbital angular momentum (OAM) mode data transmission techniques by a using a circular antenna array, operating at 28 GHz.
	The classical mode excitation of the latter is modified such that in the horizontal plane the antenna array operates as multimode  transmitter (i.e., it provides broad- , uni- and/or multi-cast transmission), while in the vertical direction OAM transmission occurs: this dual-functionality by a single  antenna-array at 28 GHz is presented for the first time.
	Specifically, it can transmit/receive in either broad-, uni-, multi-cast mode in the horizontal plane and also, it is capable of generating up to 15 spatially orthogonal OAM modes in the vertical direction. 
	The proposed circular array is designed using  twelve, low-complexity,  multi-layer microstrip patch antennas with high radiation efficiency. 
	It was tested through full electromagnetic analysis in terms of impedance matching, mutual coupling and radiation pattern.
	It was, also, fabricated and measured: good agreement between simulated and measurement results was observed.
	The proposed antenna array is perfect candidate for high spectral efficiency data transmission for 5G and beyond wireless applications. 
	%	Finally, the proposed antenna array transmits in either broad-, uni-, multi-cast mode in the horizontal plane and also, it is capable of generating up to 15 spatially orthogonal OAM modes in the vertical direction. 
	%
\end{abstract}

%	\textit{\textbf{Keywords}:}
%	Nyquist pulses, intersymbol interference (ISI), matched filters, pulse shaping methods, timing jitter.

\section*{Introduction}
\label{sec:introduction}

The  development of wireless multimedia communication techniques over the last years requires the design of new, high-directive \cite{Assimonis2018Analysis}, compact \cite{Assimonis2018Small}, reconfigurable   \cite{Assimonis2015Reconfigurable}, multiple-input-multiple-output (MIMO) \cite{Assimonis2017Analysis} and meta-surface antenna systems \cite{Assimonis2020}. 
At the same time, due to congestion in the radio frequency spectrum below $ 10 $ GHz, fifth generation of wireless communication (5G) is now exploring the best possible options at the millimeter-wave (mmWave) spectrum. 
From the wealth of literature on  communication system hardware used for the sub-$ 6 $ GHz $ 5 $G, $ 4 $G and earlier generations of cellular communication, Global System for Mobile Communications (GSM), Long Term Evolution (LTE) and wireless fidelity (Wi-Fi), we can easily envision that the most valuable resource in the future will be the efficient use of the spectrum. New disruptive technologies are currently being explored in the communication system community to preserve the spectrum as much as possible. Multimode circular array and Orbital Angular Momentum (OAM) transmitters are a few such  technologies. 
Compared to standard radio, the mode--based circular array radiation is promising in terms of spectrum efficiency \cite{Chen2018}, since it can be used for orthogonal data transmission by utilizing, at the same time, spatial and frequency resources. 

Wireless channel is inherently a broadcast medium, while uni-- and multi--cast radio employ that a conventional spatial channel is converted to beam-space channel \cite{m1,m2}. Typically, the multi--cast technique is employed by routing at the network layer. Uni-- and multi--cast techniques at physical layer means that the array equipped at the base--station (BS) or access point is creating suitable beam-patterns to serve single or multiple user groups simultaneously using same bandwidth. By doing this significant benefits in terms of spectral efficiency and reduced latency can be achieved. 
Beam-pattern shaping also means high directivity transmission of radio signal in prescribed directions, which in turn enhances the signal--to--noise--ratio (SNR) at the receiver end \cite{m3,m4}, increasing the overall communication system efficiency and throughout. Since the channel state information is only accessible at the physical layer, uni-- and multi--cast transmission at the physical layer is a complementary technique to the network-layer multi--cast routing, resulting in enhanced quality of the wireless communication system.

%In addition to multi--cast radios, several studies have investigated novel OAM mode generation approaches which uses almost the same principle of using same time and frequency resources for data transmission, but uses different radiation mechanism compared to multi--cast radiation. 
Several studies have investigated novel OAM mode radiation approaches which use the same time and frequency resources for the data transmission, while the radiation mechanism of OAM is different compared to multi--cast radiation.
Some examples of OAM radios are based on reflector-array \cite{5} metasurfaces \cite{Wang2019,4}, microstrip patch antenna \cite{4,7}, helical antenna arrays \cite{8}, and ceramic antenna arrays \cite{6}. 
Generally, a multimode array is connected to a feed network which is responsible of generating the magnitude and phase of the antenna feeds required to generate spatially orthogonal modes \cite{10}. 
OAM wave-fronts  can be generated without the requirement of feed networks  by specialized antenna geometries like in  \cite{6}. Multi--mode feed networks generally contain multiple phase shifters, switches, attenuators or radio frequency (RF) electronic devices that can control the magnitude and phase of the antenna feeding signal. For simpler cases of mode transmission, standard phase delay transmission lines are enough to realize the required antenna feed sequence. Reconfigurability is also included in some of these feed networks so that more sophisticated radio modes can be handled by the transmitter \cite{4}. Complex feed structures like the Rotman lens are also shown to be useful for OAM multimode feed networks \cite{12,13,14}.

%The contributions of this work are two--fold. First, we present high efficiency patch antenna and same antenna as a unit cell in a $ 12 $--element circular array capable of broadcast, uni-cast and multi-cast radio transmission along the entire 360$^\circ$ azimuth plane (Fig. \ref{fig1}) while the same circular array is capable of OAM mode transmission in the vertical direction. Secondly, we show that the circular array is capable of generating as many as $ 15 $ spatially orthogonal OAM modes. 
The contributions of this work is two-fold. First, we present a high efficiency $ 12 $--element circular array operating at $ 28 $ GHz, capable of broadcast, uni--cast and multi--cast radio transmission along the entire $ 360 $ deg. azimuth plane. 
%
%while the same circular array is capable of OAM mode transmission in the vertical direction. 
%
Secondly, we show that the same circular array is capable of generating as many as $ 15 $ spatially orthogonal OAM modes in the vertical direction: this dual-functionality by a single antenna-array at $ 28 $ GHz is presented for the first time in the literature. The antenna array was tested through full electromagnetic analysis and measurements in terms of reflection coefficients, mutual coupling between the array elements and radiation pattern (near-- and far--field).
Particularly, Section II of the paper presents the synthesis technique and results of the circular array, section III discusses the uni-- or multi--cast capabilities, section IV presents the spatially orthogonal OAM mode generation, while the study is concluded in section V of the paper. For the rest of the paper, circular array modes \cite{circ} along azimuth plane are represented by $m$ while the OAM modes along the elevation direction are represented by $\ell$.

\section*{Results and Discussion}
\subsection*{Twofold Function Circular Antenna Array}

In this section the proposed antenna will be theoretically analysed and designed though full-electromagnetic analysis. All results will be cross-check through measurements.

\subsubsection*{Theoretical Analysis}

The radiation patterns of a circular antenna array with $ N $ equally spaced elements and radius of $ R $, which is placed in the horizontal plane and specifically it centre lies  at the origin of coordinates, is given in the classical array literature \cite{van2004optimum,Sheleg1968Matrix}, 
\begin{equation}\label{radiation_pattern_CA}
	E^{\,m} \left( \theta, \phi \right) = \sum \limits_{n=0}^{N-1} \mathbf{w}^{\,m} E_n \, e^{j k R \sin \theta \, \cos\left( \phi -\phi_n \right) }
\end{equation}
where,
\begin{equation}\label{radiation_pattern_phi_n}
	\phi_n = n \frac{2 \pi}{N},
\end{equation}
is the inter--element phase--shift,
$ \theta, \phi $ is the polar, azimuthal angle respectively,
$ k=2\pi/\lambda $ is the wavenumber and $ \lambda $ is the wavelength at the operating frequency,
$ E_n $ is the single antenna element's radiation pattern, which is usually assumed that $ E_n =1$, indicating isotropic elements,
and 
\begin{equation}\label{current_excitation}
	%\mathbf{w}_m = \left[1,~e^{j n m},~e^{j 2 n m},~\cdots~,~e^{j \left(N-1 \right)  n m} \right]^{T}
	\mathbf{w}^{\,m} = \left[\cdots,~e^{j m  \phi_n},~\cdots\right]^{T}.
\end{equation}
is the  vector related to the $ m- $th mode, which represents the current excitation on each element, where, 
\begin{equation}\label{m}
	m = 0,~\pm 1,~\cdots,~\pm\left( \frac{N}{2}-1\right) ,~\frac{N}{2} ,
\end{equation}
assuming that $ N $ is even number, without loss of generality. Thus, for an antenna array with $ N$ elements it is possible to excite $ N $ modes. 
In order to steer the beam to a specific direction  along the horizontal plane, an extra angle, $ \psi $, can be utilized as follows:
\begin{equation}\label{current_excitation2}
	%\mathbf{w}_m = \left[1,~e^{j n m},~e^{j 2 n m},~\cdots~,~e^{j \left(N-1 \right)  n m} \right]^{T}
	\mathbf{w}^{\,m} = \left[\cdots,~e^{j m  \left( \phi_n + \psi\right) },~\cdots\right]^{T}.
\end{equation}
Specifically,  beam rotates clockwise by angle $ \psi $.

In general, the circular array  mode zero ($ m = 0 $) gives an omnidirectional radiation pattern (\textit{broadcast}), albeit with low directivity. The sum of all the modes results in a directional radiation pattern, with one main lobe (\textit{uni--cast}), but now with higher directivity. 
%
%It is noted that, usually, the use of higher modes for uni--cast transmission leads to poorest results in terms directivity \cite{Sheleg1968Matrix}. 
%
Multi--mode circular antenna array also allows \textit{multi--cast} transmission: by summing  all the modes expect  zero mode, the final radiation pattern includes multiple lobes, and thus, antenna  can transmit to different directions.

\begin{figure}[!t]
	\centering
	\includegraphics[width=0.5\linewidth]{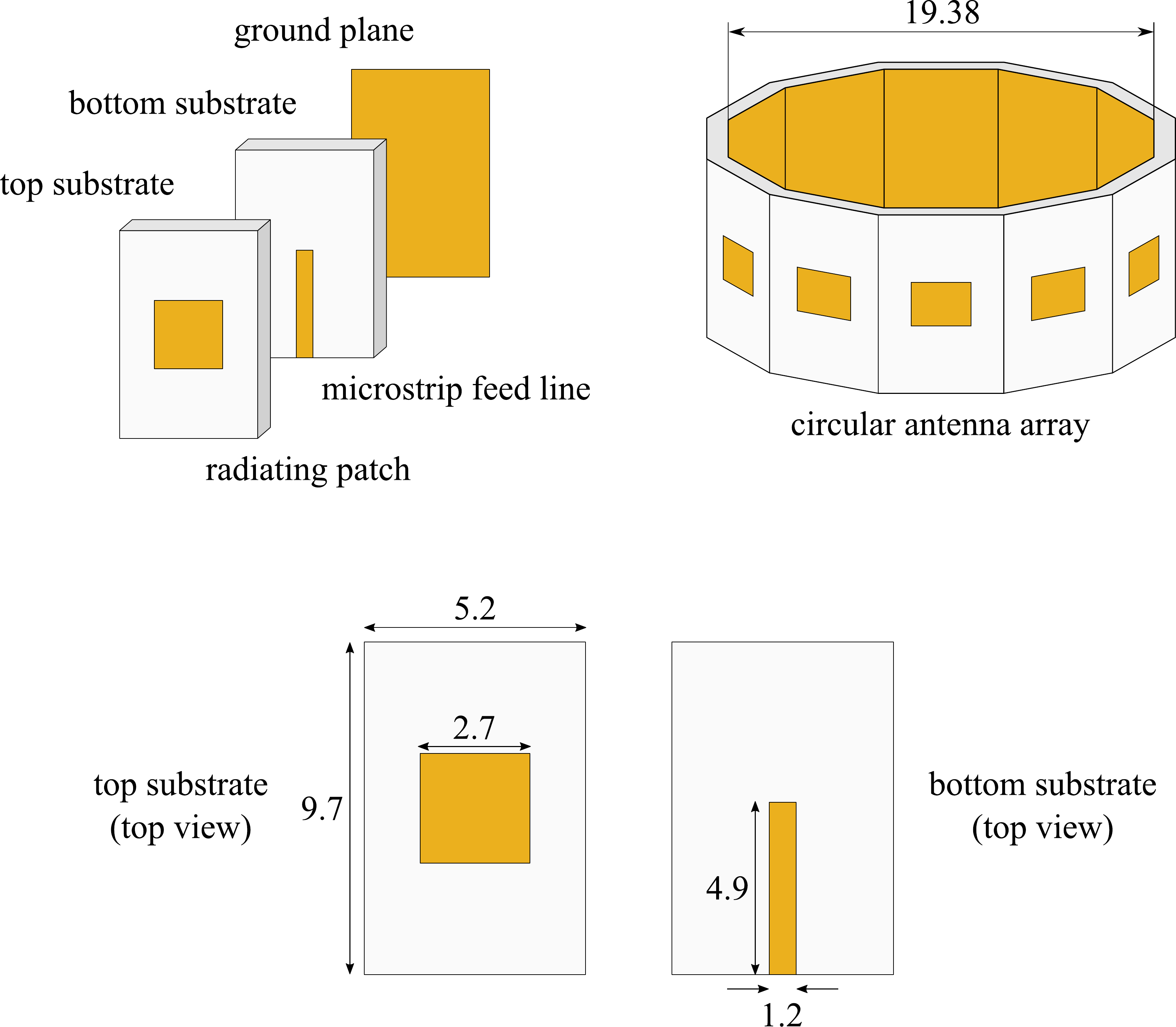}
	\caption{ Structural configuration multilayer (two substrates) microstrip patch antenna and the $ 12 $-element circular patch antenna array illustration. Also depicted the dimensions of the antenna element. Radiating rectangular patch with edge of $2.7 $ mm is placed in the centre of the top substrate of dimensions $ 5.2~\mathrm{mm} \times 9.7 ~\mathrm{mm}  \times 0.5 ~\mathrm{mm}  $. The bottom substrate has the same dimensions with the top and here the microstrip feed line with dimensions $ 4.9~\mathrm{mm} \times 1.2 ~\mathrm{mm}   $ is placed.}
	\label{fig1}
\end{figure}

\begin{figure}[t!]
	\centering
	\subfloat[]{
		\includegraphics[width=0.33\linewidth]{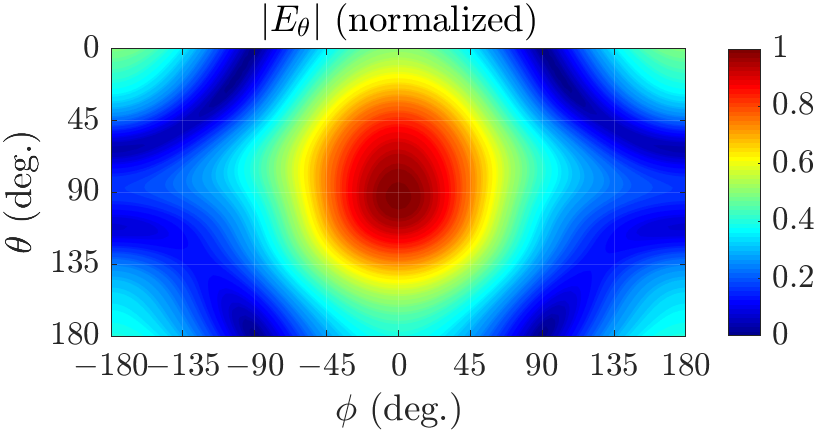}
		\label{fig2A}
	}
	\hfil
	%	\\
	\subfloat[]{
		\includegraphics[width=0.33\linewidth]{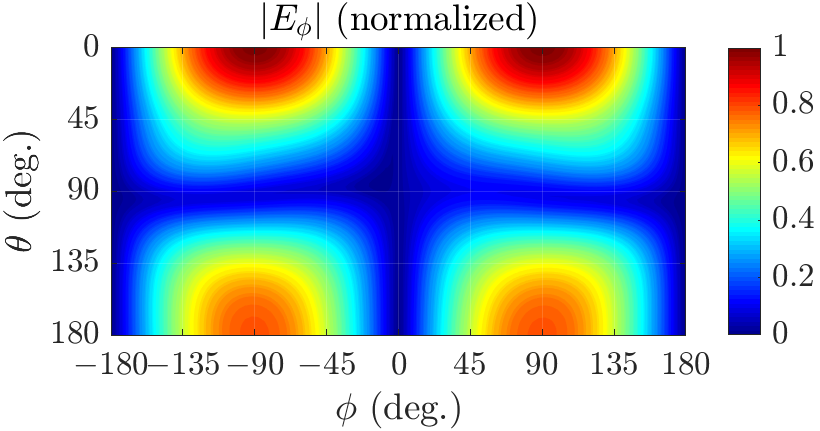}
		\label{fig2B}
	}
	\caption{Simulated radiation pattern (a) $\theta$, and (b) $\phi$ component of the \textit{E}-field. 
		In the horizontal plane (i.e., $ \theta = 90 $ deg.) the $ E_{\theta} $ component is maximized at $ \phi=0 $ deg., allowing multimodal operation, as will explained, whilst in the vertical direction, at the top of the circular antenna array, $ E_{\phi} $ component is maximized at $ \phi=\pm{}90 $ deg., i.e., where the $ E_\theta $ vanishes. Actually, where $ E_\phi $ is maximized $ E_\theta $ is minimized and vice-versa.
		The latter is the key-point that makes the proposed circular antenna  twofold functional.}
	\label{fig2}
\end{figure}

\begin{figure}[!t!]
	\centering
	\subfloat[]{
		\includegraphics[width=0.33\linewidth]{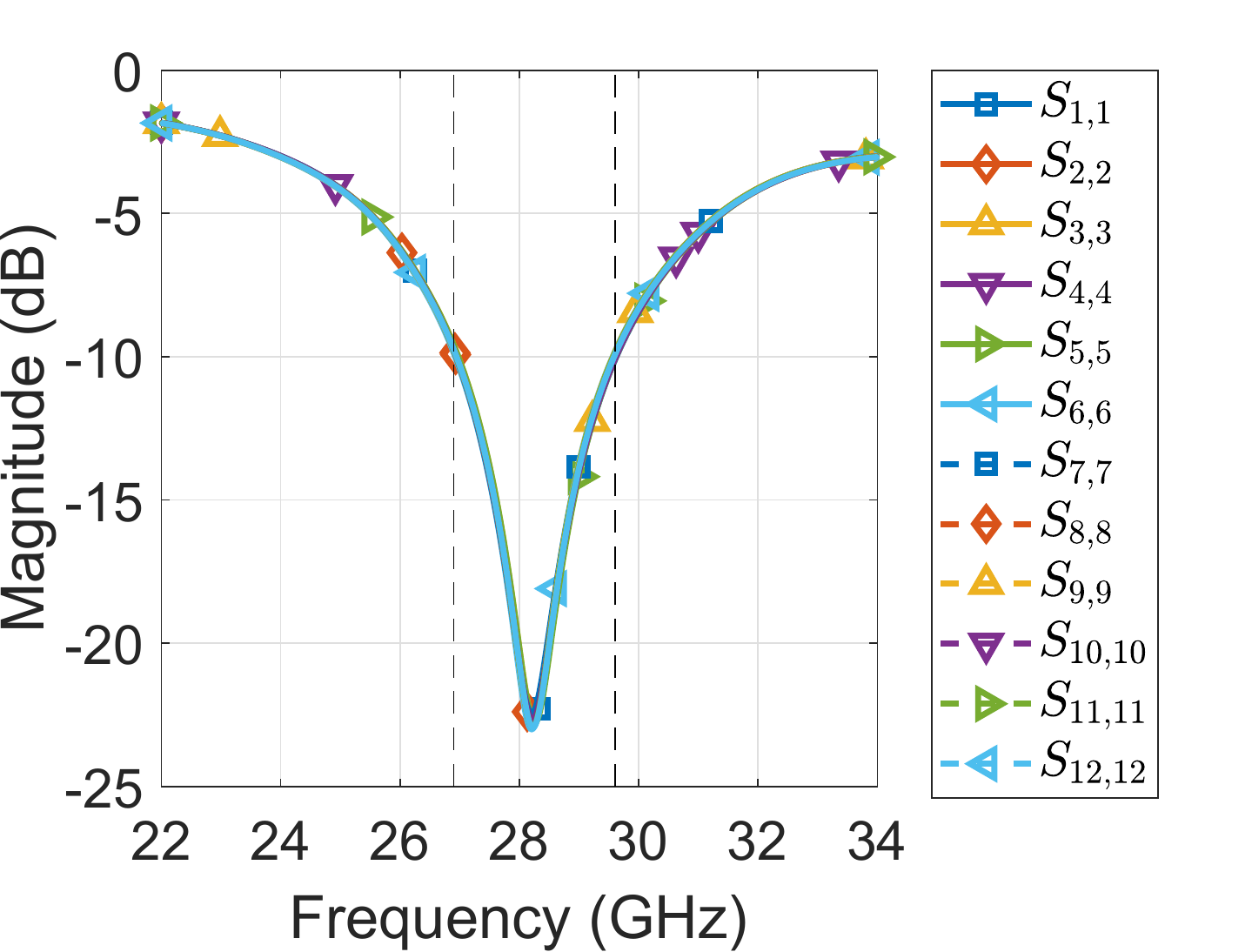}
		\label{fig3A}
	}
	%	\\
	\hfil
	\subfloat[]{
		\includegraphics[width=0.33\linewidth]{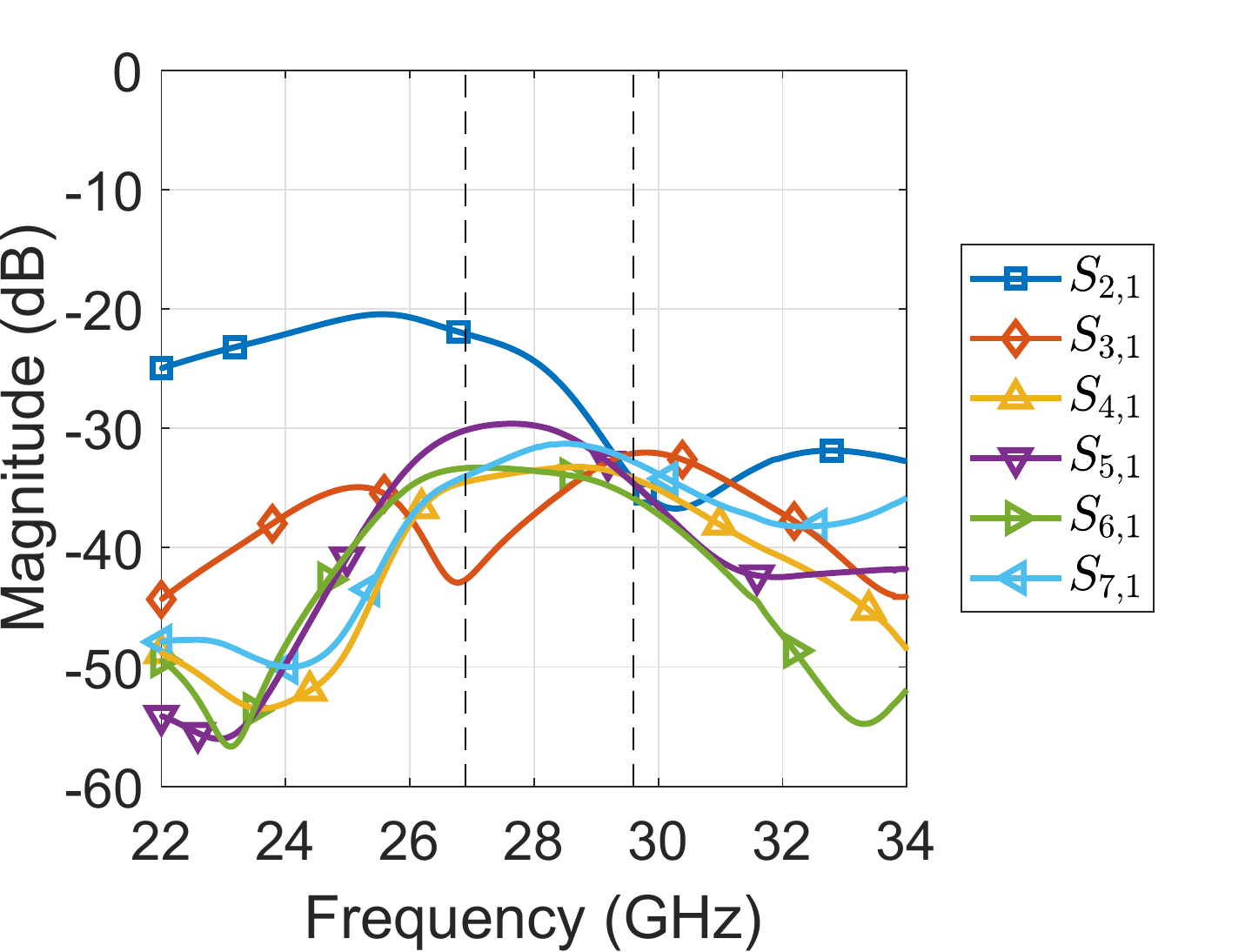}
		\label{fig3B}
	}
	\caption{Simulated $S$-parameters of $ 12 $-element patch antenna array at the input ports indicating (a) the return loss and (b) the cross coupling.}
	\label{fig3}
\end{figure}
\begin{figure*}[!t!]
	\centering
	\subfloat[]{
		\includegraphics[width=0.225\linewidth,valign=m]{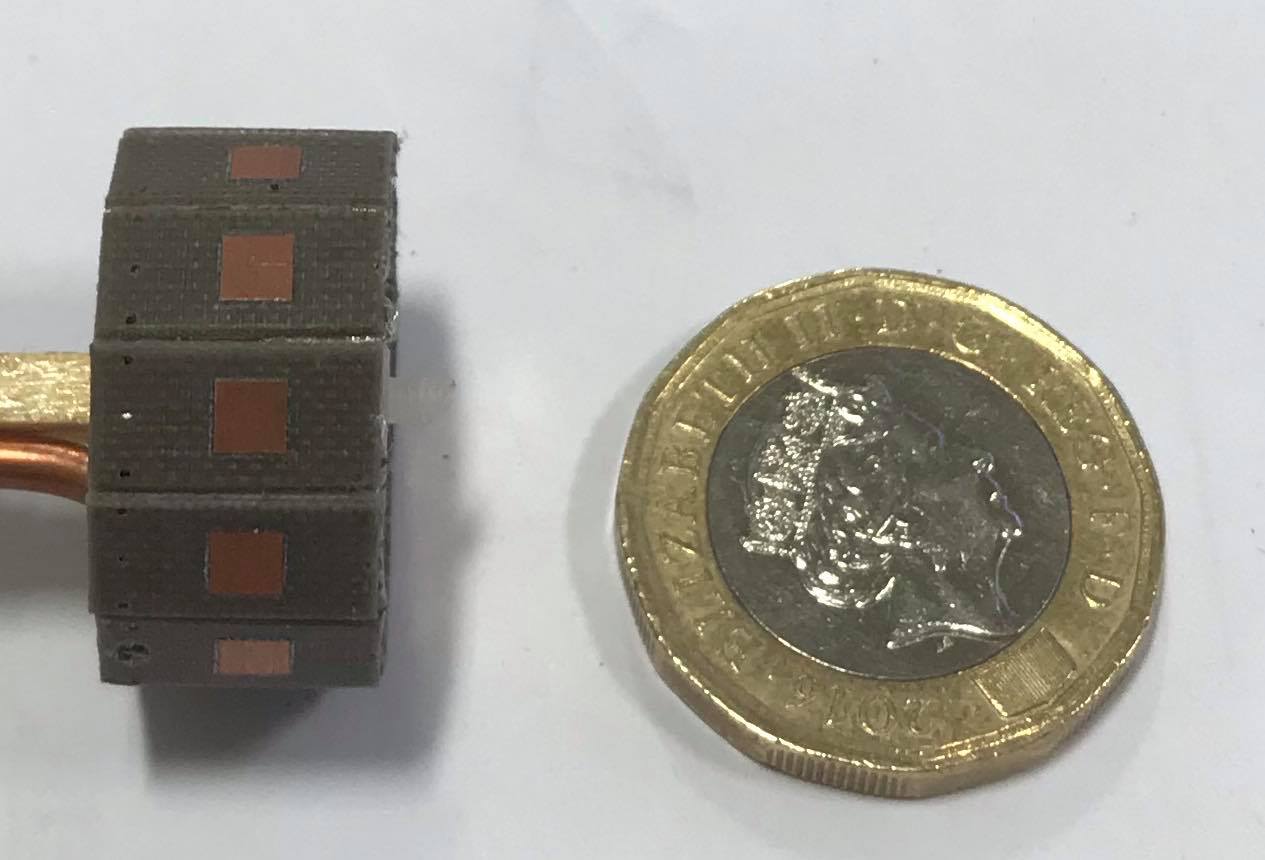}
		\vphantom{
			\includegraphics[height=0.23\linewidth,valign=m]{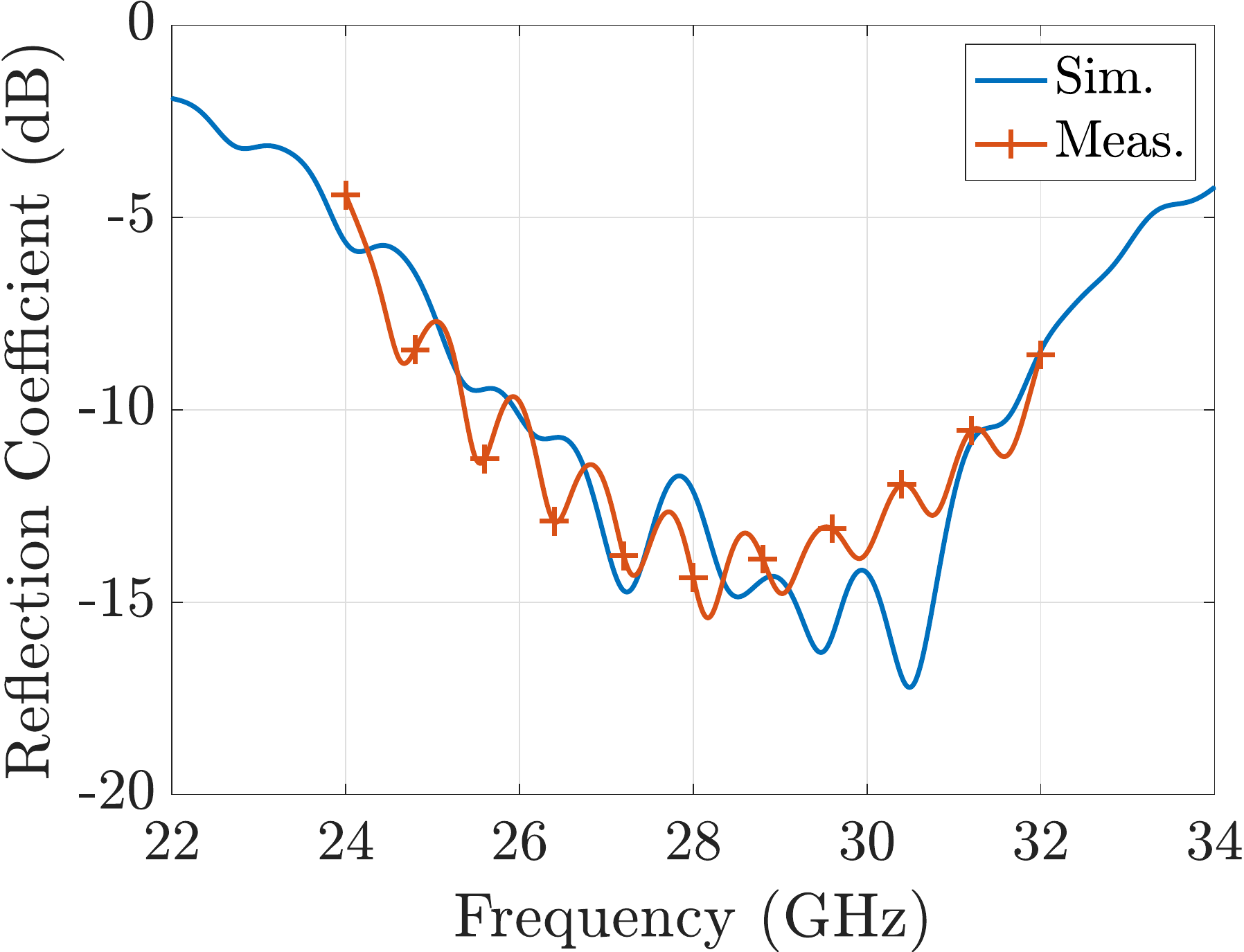}
		}
		\label{fig4A}
	}
	%	%	\\
	%	\hfil
	%	\subfloat[]{
	%		\includegraphics[height=0.19\linewidth,valign=m]{Fig004B.PNG}
	%		\vphantom{
	%			\includegraphics[height=0.23\linewidth,valign=m]{Fig004C.pdf}
	%		}
	%		\label{fig4B}
	%	}
	%	\\
	\hfil
	\subfloat[]{
		\includegraphics[height=0.23\linewidth,valign=m]{Fig004C.pdf}
		\label{fig4C}
	}
	\\
	%	\hfil
	\subfloat[co--polarized]{
		\includegraphics[width=0.47\linewidth,valign=m]{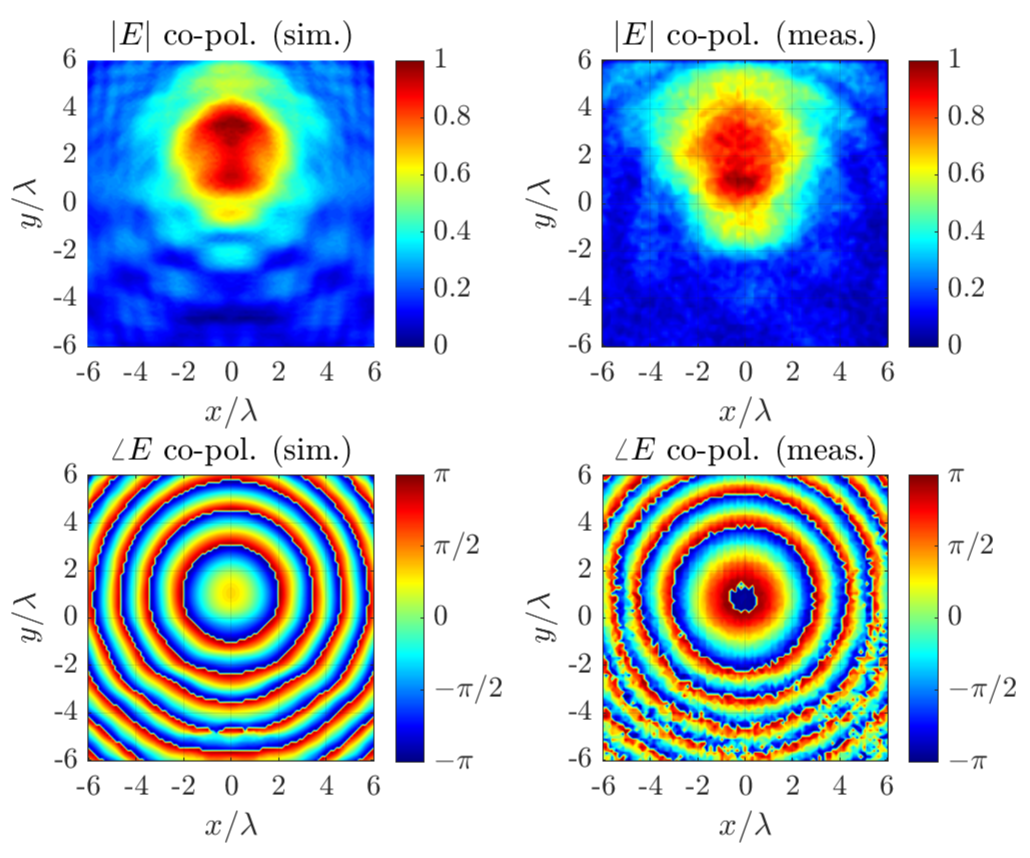}
		\label{fig4D}
	}
	\hfil
	\subfloat[cross--polarized]{
		\includegraphics[width=0.47\linewidth,valign=m]{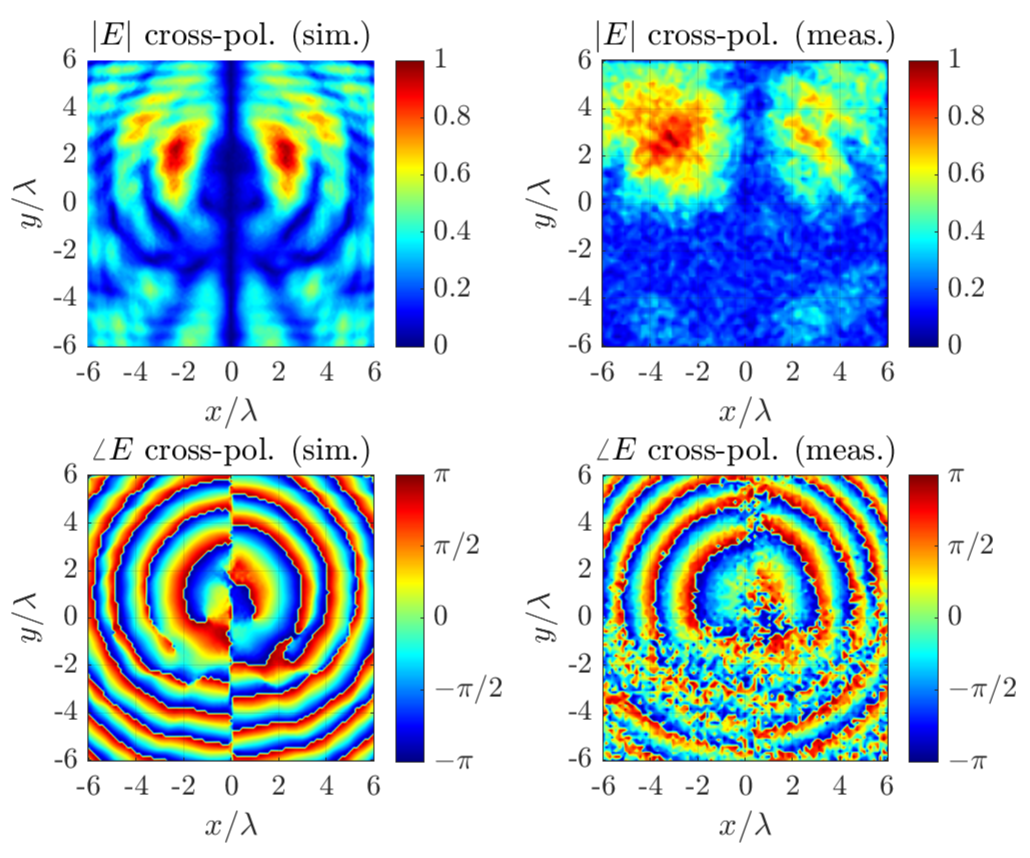}
		\label{fig4E}
	}
	\caption{ (a) The fabricated and tested circular antenna array and (b) the measurement set-up. 
		(c) Comparison between simulated and measured return loss at the end of the coax feed connected to one antenna element when all other elements' ports are open-circuited. 
		Comparison between simulated and measured $E$-field patterns in the form of magnitude and phase for (d) Co- and (c) Cross-polarized component patterns: a good agreement is observed.}
	\label{fig4}
\end{figure*}

\subsubsection*{Circular Patch Antenna Array Design}

To realize the circular array, we use an electromagnetically coupled patch antenna designed using two substrate layers. The bottom layer is designed on Rogers RO4003 substrate ($\epsilon_r = 3.38$, $\tan\delta = 0.0027$) while the top layer is designed on Taconic TLY--5A substrate ($\epsilon_r = 2.17$, $\tan\delta = 0.0009$). The antenna element is fed by a micro-strip transmission line \cite{wong2004compact}, designed on Taconic TLY--5A with full ground plane. The microstrip line is fed by a $ 50 $ $\Omega$ lumped port when the antenna is simulated in full wave electromagnetic simulation tool CST Microwave Studio. The patch antenna geometry and optimized design parameters are presented in Fig. \ref{fig1} when the antenna is placed in $xy$--plane. Here, it can be observed that the microstrip patch is fed at the centre such that the patch radiation is isolated from the feed line and there is a gap between the feeding mechanism and radiating patch. 
This allows high directivity radiation along the $+z$-- as well as $+y$--directions with realized gain of $ 5.74 $ dBi and consistent half--power beam width of $ 100 $ deg. and $ 104 $ deg. along $xz$- and $yz$-planes. The benefit of this antenna radiation mechanism will be shown in the later section when same antenna will be used for multi-cast and OAM mode radiation in different directions. The antenna is matched to operate at $ 28 $ GHz with $ 98 $$\%$ simulated radiation efficiency.

We can see from Fig. \ref{fig2A} and  \ref{fig2B} that in the horizontal plane ($\theta = 90$ deg.) the maximum $E$-field occurs for $\phi$ = 0 deg. 
At this plane, $E$ is equal to the component  $E_\theta$, which is vertical to the horizontal plane, while component $E_\phi$ is zero.
On the other hand, $E_\phi$ is maximized at $\theta = 0$ deg., i.e., at the top of the antenna, and for $\phi = \pm 90$ deg., where $E_\theta$ vanishes.
Actually, where $E_\phi$ is maximized $E_\theta$ is minimized and vice-versa. Thus, in the horizontal plane, the dominant electric component in the far-field is the $E_\theta$, which is perpendicular to the horizontal plane and leads to the multimodal function, while in the vertical direction, on the top of the antenna, the dominant electric component  in the far-field is the $E_\phi$, which is parallel to the horizontal plane and leads to the OAM function.

Copies of the patch antennas are placed in a circular array formation same as in Fig. \ref{fig1} in (horizontal) $xy$--plane when the array diameter $d = 19.38$ mm ($ 1.8\lambda $). 
%The array antennas are then re-optimized concurrently to operate at $ 28 $ GHz. 
%Simulated results show each antenna operation from $ 26.95 $ to $ 29.64 $ GHz (9.5$\%$ bandwidth) with a return loss lower than $ -10 $ dB. 
%
Simulated $S$--parameters of the circular array  is shown in Fig. \ref{fig3}, when all other antenna ports are terminated to $ 50 $ $\Omega$ matched load to test simultaneous array excitation. 
Reflection coefficient results (Fig. \ref{fig3A}) show each antenna operates from $ 26.95 $ to $ 29.64 $ GHz (9.5$\%$ bandwidth) with a return loss lower than $ -10 $ dB. 
Due to cylindrical symmetry of the array based on \eqref{radiation_pattern_CA}, cross coupling results against excitation of only $ 7 $ ports are shown in Fig. \ref{fig3B}. The worst cross coupling is between immediate neighbouring elements ($ -24 $ dB at $ 28 $ GHz) while it is below this value for the rest of the cases.

The array is realized using a combination of separately fabricated patch antennas, each connected to a coaxial feed line as shown in Fig. \ref{fig4A}. The array antenna matching is measured by first connecting single antenna coaxial feed to a  Vector Network Analyser while all other antenna ports are kept open. The return loss is shown in Fig. \ref{fig4C}. After confirmation of good matching, the antenna array is placed in a planar near--field measurement facility and magnitude and phase responses are recorded along the $+z$--direction. Measurements were taken using vertically and horizontally polarized measurement probe when circular array is placed in the near field of the 28 GHz probe. Results shown in \ref{fig4D} and \ref{fig4E} verifies the simulated predictions; a clear correlation between measured and simulated fields can be observed from the magnitude and phase plots for both co-- and cross--polarization fields. After successful verification of the circular array operation same array is used to generate multi--cast and OAM modes that are discussed in the proceeding sections.

\subsection*{Uni-- and Multi--cast Transmission}

\begin{figure*}[!t!]
	\centering
	\subfloat[broadcast]{
		\includegraphics[width=0.25\linewidth,valign=m]{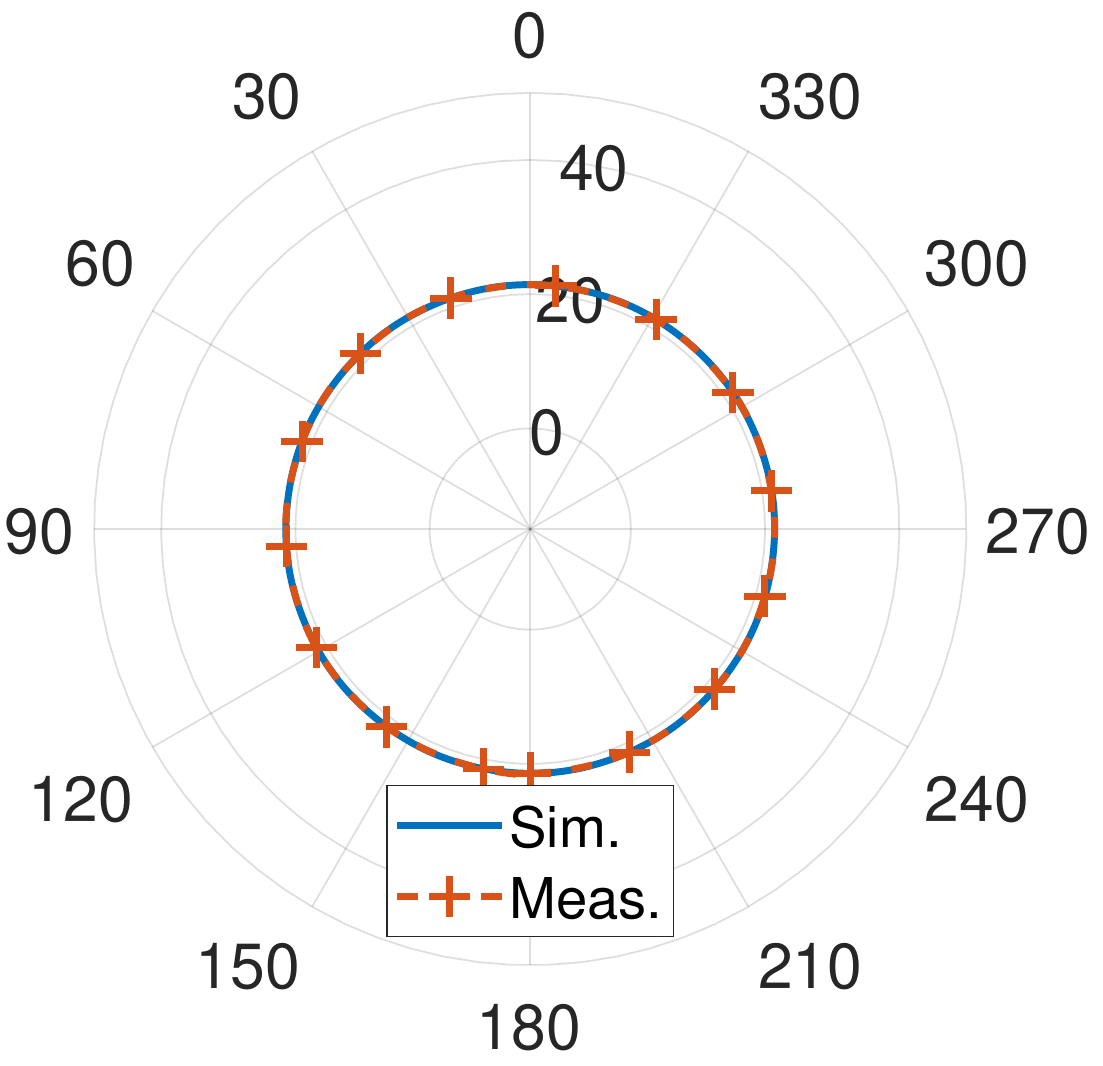}
		%		\vphantom{
		%			\includegraphics[width=0.3\linewidth,valign=m]{u1t.pdf}
		%		}
		\label{fig5A}
	}
	\hfil
	\subfloat[uni--cast \textit{A}]{
		\includegraphics[width=0.25\linewidth,valign=m]{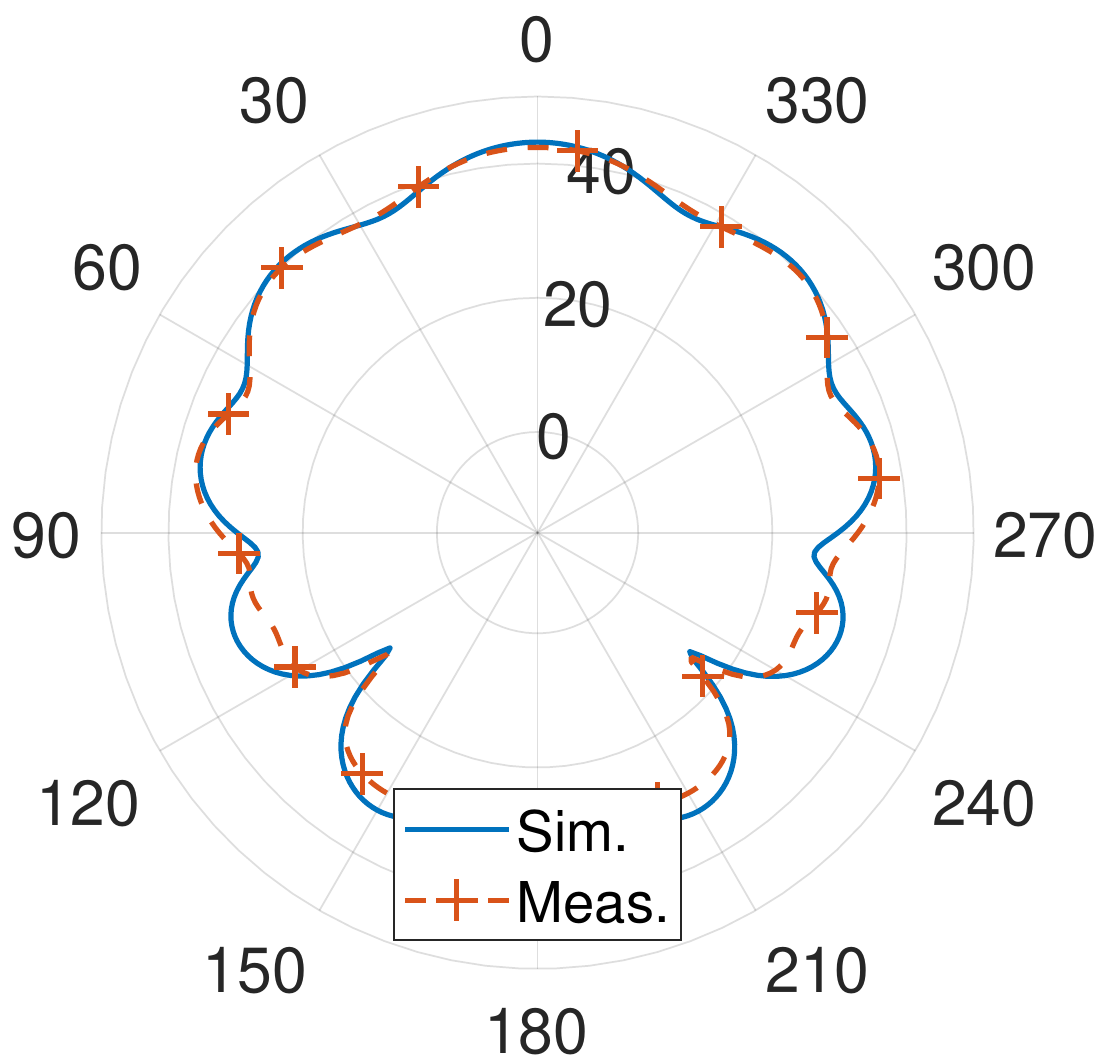}
		%		\vphantom{
		%			\includegraphics[width=0.3\linewidth,valign=m]{Fig004B.pdf}
		%		}
		\label{fig5B}
	}
	\hfil
	\subfloat[uni--cast \textit{A}]{
		\includegraphics[width=0.25\linewidth,valign=m]{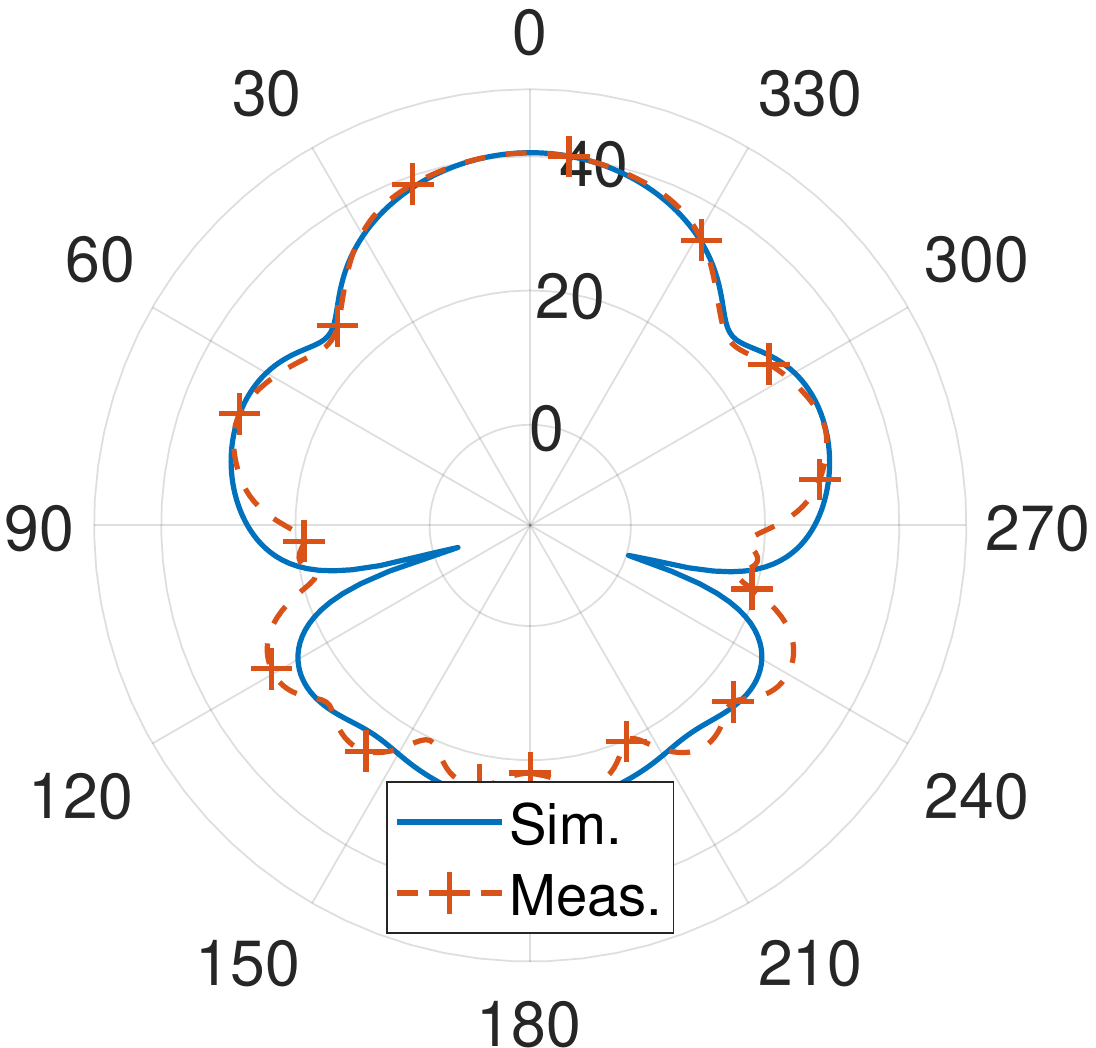}
		\label{fig5C}
	}
	\\
	%	\hfil
	\subfloat[multi--cast \textit{A}]{
		\includegraphics[width=0.25\linewidth,valign=m]{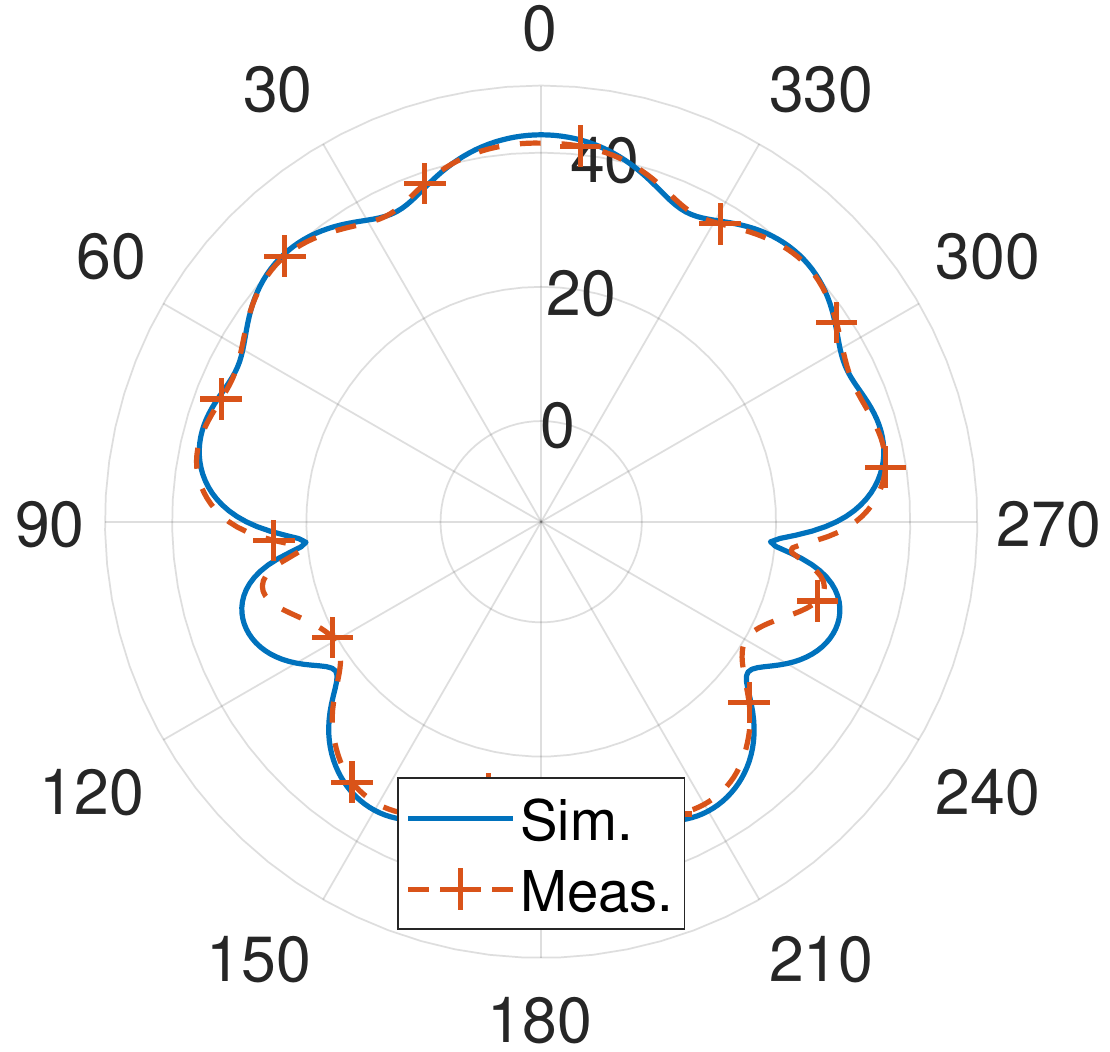}
		\label{fig5D}
	}
	\hfil
	\subfloat[multi--cast \textit{B}]{
		\includegraphics[width=0.25\linewidth,valign=m]{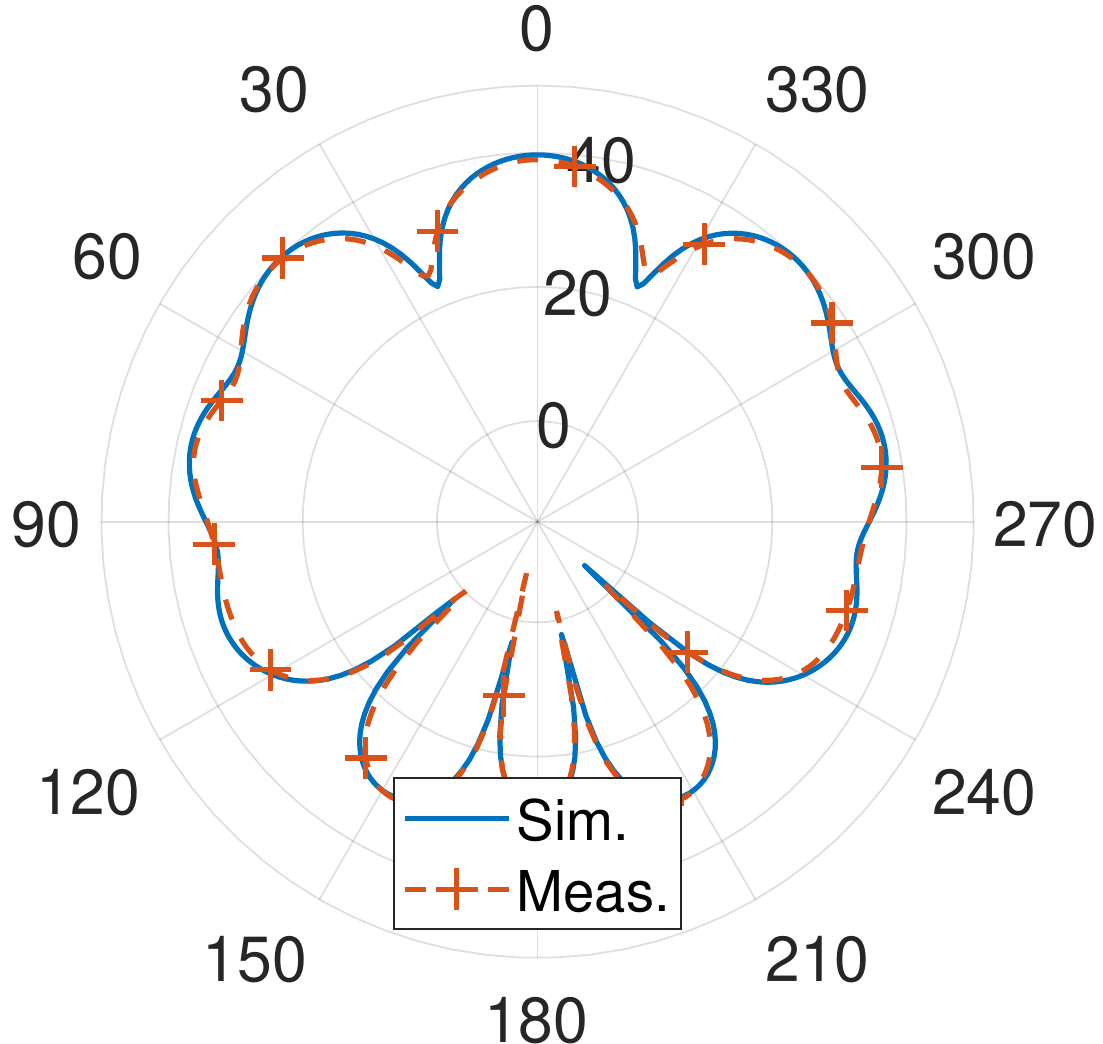}
		\label{fig5E}
	}
	\hfil
	\subfloat[multi--cast \textit{C}]{
		\includegraphics[width=0.25\linewidth,valign=m]{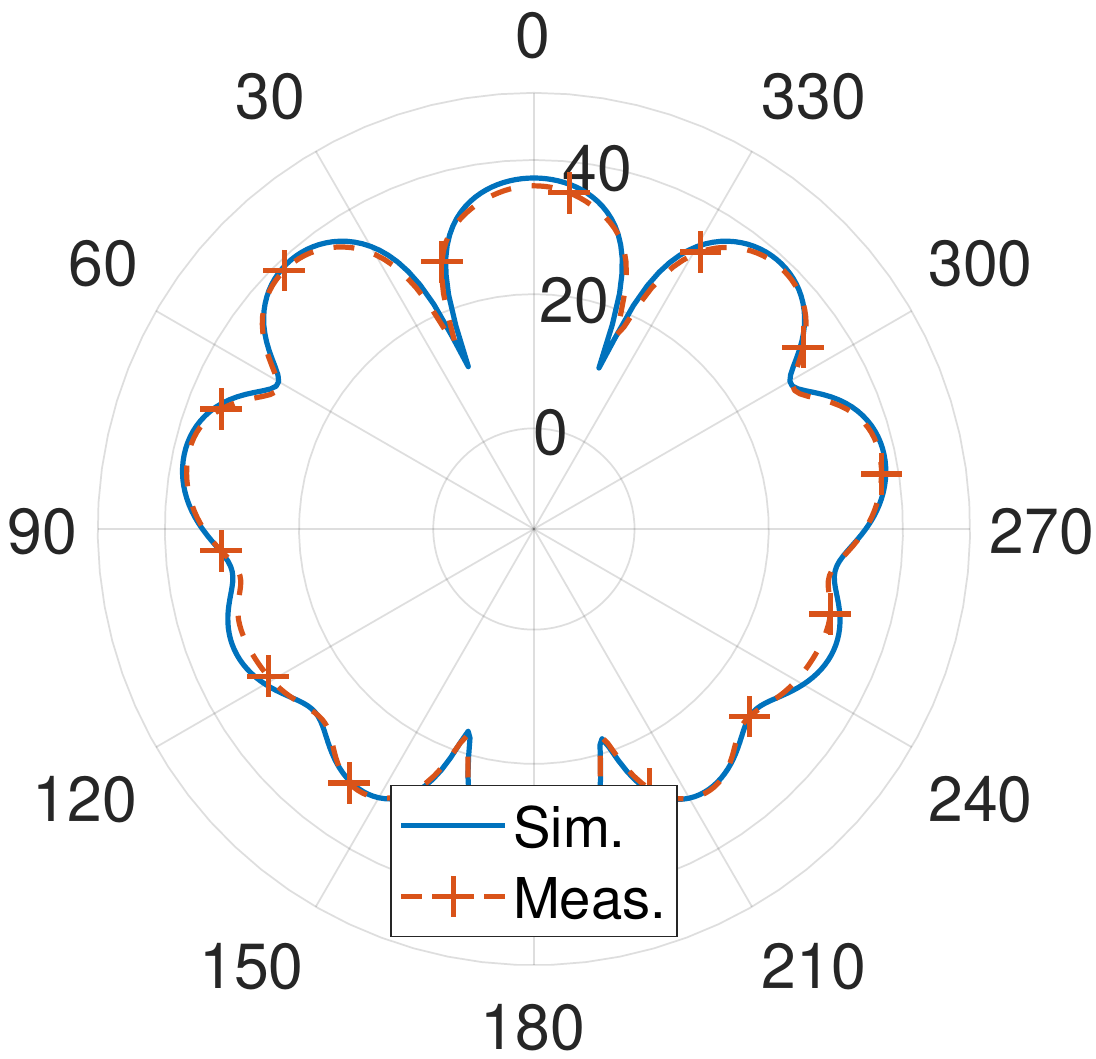}
		\label{fig5F}
	}
	\caption{ Comparison between simulated and measured electric far-field patterns in dB scaling for (a) general broadcast, (b) uni–cast \textit{A}, (c) uni–cast \textit{B}, (d) multi–cast \textit{A}, (e) multi–cast \textit{B}, and (f) multi–cast \textit{C}: a good agreement is observed.}
	\label{fig5}
\end{figure*}

The mode-mixing excitation of a circular array allows uni-cast and multi-cast radio transmission in addition to broadcast, and this is done by using only some of the $m$ modes from \eqref{m}. 
In this section, we show that the patch antenna circular array can generate two uni--cast and three multi--cast radio transmissions. 
Two cases of uni-cast \textit{A} and \textit{B} are shown which are resulted by the mode mixing $m = 0, \pm1, ...\pm5$ and $m = 0, \pm1, \pm 2, \pm3$. Multi-cast has three cases; \textit{A}, \textit{B} and \textit{C}. The case \textit{A} is a result of mode mixing of $m = \pm1, \pm2, ..., \pm5$, \textit{B} is a result of mode mixing $m = \pm3, \pm4, \pm5$, and for \textit{C} it is $m = \pm4, \pm5$. 
The simulated results for the electric component in the far--field, estimated  thought CST Microwave Studio is depicted  in Fig. \ref{fig5}.
First, the broadcast radiation pattern ($ m=0 $) in the horizontal plane, where all the elements are exited by the same signal and the phase difference between them is zero, is depicted in Fig. \ref{fig5A}: the electric field is omni--directional, as expected in that case, with a maximum of $ 22 $ dBV/m and corresponding directivity of $ 0 $ dBi:
it is noted that the maximum directivity in that case is $ 4.86 $ dBi but occurs in the plane $ \theta=15 $ deg. and not in the horizontal plane (i.e., $ \theta=90 $ deg.)
Electric field for the uni--cast modes \textit{A} and \textit{B} is depicted in Fig. \ref{fig5B} and \ref{fig5C} and has maxima of $ 44.5 $  and  $ 42.8 $ dBV/m, respectively, with corresponding directivity of $ 8.59 $ and $ 8.91 $ dBi.
%, compared to a general broadcast directivity \cite{circ} of $ 4.86 $ dBi (Fig. \ref{fig4A}). 
%
For the multi--cast cases, simulated results are shown in Fig. \ref{fig5D}--\ref{fig5F}, where the circular array is capable of data transmission in three dominant directions, i.e., $ 0 $ deg., $ \pm{}45 $ deg. Here the corresponding peak directivity at $ 0 $ deg. decreases from $ 8.1 $ dBi (multi--cast case \textit{A}) to $ 7.7 $ dBi (case \textit{B}) to $ 6.6 $ dBi (case \textit{C}) at an expense of the radiated power from the array in multiple directions being equalized. This is evident from Fig. \ref{fig5F} where almost equal power is radiated along three directions.

%\begin{figure*}[!t!]
%	
%	\centering
%	\subfloat[$ \ell =0 $]{
%		\includegraphics[width=0.475\linewidth,valign=m]{Fig006A.PNG}
%		%		\vphantom{
%		%			\includegraphics[width=0.3\linewidth,valign=m]{u1t.pdf}
%		%		}
%		\label{fig6A}
%	}
%	\hfil
%	\subfloat[$ \ell = \pm1 $]{
%		\includegraphics[width=0.475\linewidth,valign=m]{Fig006B.PNG}
%		%		\vphantom{
%		%			\includegraphics[width=0.3\linewidth,valign=m]{Fig004B.pdf}
%		%		}
%		\label{fig6B}
%	}
%	\caption{ Predicted normalized far--field directivity patterns of OAM modes (a) $\ell = 0$, (b) $\ell = \pm1$, and (c) $\ell = \pm2$, (d) $\ell = \pm3$, (e) $\ell = \pm4$, (f) $\ell = \pm5$, (g) $\ell = \pm6$, and (h) $\ell = \pm7$. 
%		%
%		The phase/magnitude colour map scales and axis shown in (a) are common for all the OAM modes.}
%	\label{fig6}
%\end{figure*}

Broadcast and all scenarios of uni--cast and multi--cast are cross-verified with the measurements taken in anechoic environment. The circular array is placed in a turn table of the far-field anechoic chamber, as will be explained in Measurements section,
and its radiated electric field in the horizontal plane is measured along $ 360 $ deg. through a receiver   horn antenna \cite{horn} operating at $ 28 $ GHz and also depicted in Fig. \ref{fig5} for comparison. 
Only a singe antenna element of the circular antenna array is fed   and the total depicted measured radiation patterns are then mathematical estimated based on basic antenna array theory \cite{Sheleg1968Matrix}: the total field is the superposition of the single element's pattern because of the antenna array's cylindrical symmetry.
Good agreement between simulated, measured results is observed.
%
%Based on \eqref{label} note that, for all the case shown in Fig. \ref{fig5}, the radiated field is rotatable along $ 360 $ deg. along the horizontal-plane by adding an equal phase delay in all the antenna excitations/elements. 
%
\begin{figure*}[!t]
	\centering
	%	\subfloat[]{
	\includegraphics[width=0.33\linewidth,valign=m]{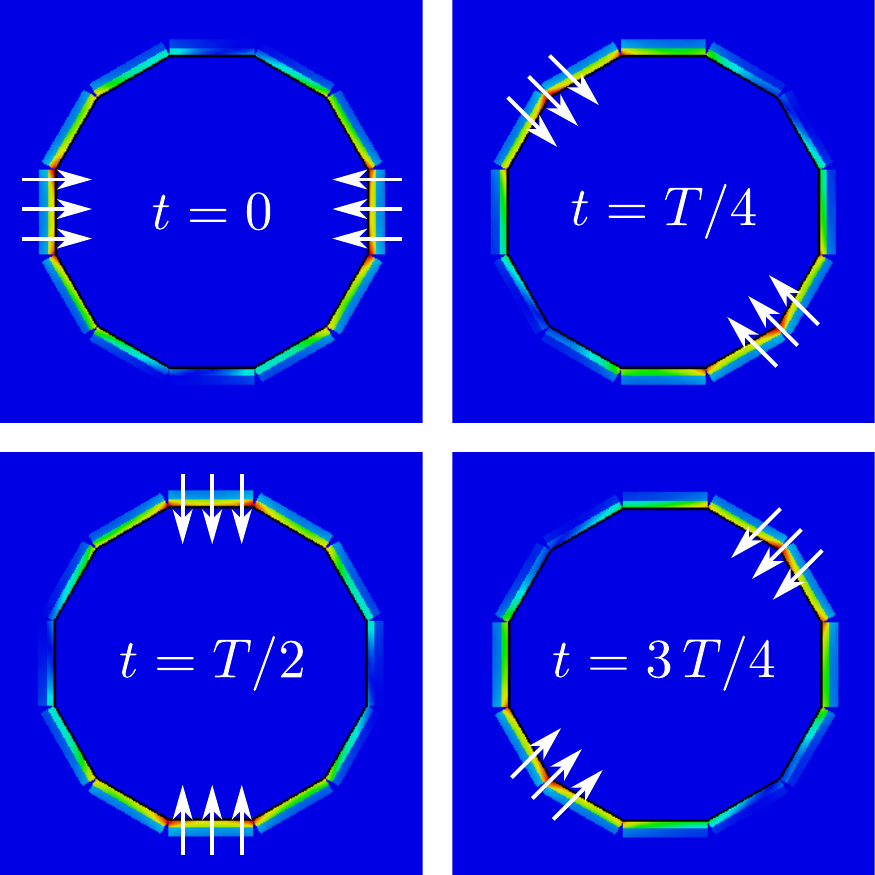}
	%		\label{fig7A}
	%	}
	%	\hfil
	%	\subfloat[]{
	%		\includegraphics[width=0.475\linewidth,valign=m]{Fig007B.png}
	%		\label{fig7B}
	%	}
	\caption{Simulated current distribution of the antenna array at different time phases, where $ T $ is the period, for the first mode (i.e., $ \ell = +1 $) at $ 28 $ GHz.}
	\label{fig78}
\end{figure*}
These results are considered as reference to compare the performance of uni-- and multi--cast transmission. Uni--cast case \textit{B} seems to be the highest directivity transmission option, whilst multi--cast case \textit{B} and \textit{C} can be good choices when simultaneous data transmission is desirable along three directions around the circular array. 

Also, please note that, all the beams for the uni-- and multi--cast transmission (Fig. \ref{fig5})) can be rotated by angle $ \psi $, based on  \eqref{current_excitation2}, and thus, the proposed, multi-mode antenna is  \textit{steerable}, providing $ 360 $ deg. rotation in the horizontal plane. 
%
%The angle $ \psi $ can be implemented by applying
In order to dynamically change the current's phase on each antenna element, various practical techniques can be applied, e.g., Rotman Lens and phase ramps \cite{12,13,14}.

\subsection*{OAM Mode Transmission}

Advantageously, the same circular antenna array in Fig. \ref{fig1} and \ref{fig4A} can generate up to $ 15 $ OAM modes along $\pm z$--direction (i.e., $ \theta = 0 $ deg.) when the array is placed in the reference horizontal plane ($ xy $--plane). 
The generation of OAM modes is a consequence of radiation characteristics of the single antenna element, as previously described.
Based on Fig.  \ref{fig2} it is evident that the antenna is capable of transmitting in both ways, well separated from each other.
Specifically, in the horizontal plane the dominant electric component is the $ E_\theta $, resulting vertical polarization, and thus, multimodal operation, whilst, in the vertical direction the dominant component is the $ E_\phi $, which in tern, is coupled with resulting radiated field of the other elements of the array, leading to OAM operation.
Putting this statement into perspective, note the gap between feeding line on Rogers RO4003 substrate and radiating patch on Taconic TLY-5A substrate in Fig. \ref{fig1}. When the radiating patches are placed in circular array formation, and if we observe the array from the top, we can observe a series of radiating copper edges tangential to the array circle (Fig. \ref{fig78}). These edges act as resonant slots and is primarily responsible for majority of the OAM modes. Mathematically, the excitation vector for the $N$ element circular array is given by
\begin{equation}\label{current_excitation_ell}
	w_i = e^{2j(i-1){\ell \pi}/{N}},
\end{equation}
where, $i = 1,2,...N$ represents the antenna elements of the array and $\ell$ is the OAM mode number. 
The number of the antenna elements in the circular array determines the number of the OAM modes that can be produced \cite{PhysRevLett.99.087701}. 
The radiation mechanism, which leads to OAM transmission is presented in Fig. \ref{fig78}. Specifically, it is depicted the simulated surface current distribution of the antenna array at different time phases ($ T $ is the period) for the first mode ($ \ell=+1 $). As the period increases by $ 45 $ deg., the current phase is shifting accordingly. Also, the current vectors have opposite direction, as described in \cite{PhysRevLett.99.087701} for the same mode ($ \ell=+1 $), which leads to OAM transmission.

\begin{figure*}[!t!]
	
	\centering
	\subfloat[$ \ell =0 $]{
		\includegraphics[width=0.475\linewidth,valign=m]{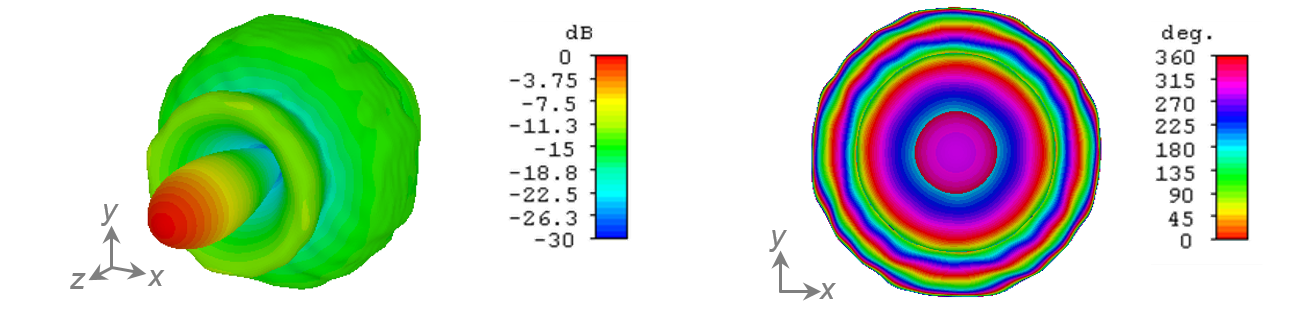}
		%		\vphantom{
		%			\includegraphics[width=0.3\linewidth,valign=m]{u1t.pdf}
		%		}
		\label{fig6A}
	}
	\hfil
	\subfloat[$ \ell = \pm1 $]{
		\includegraphics[width=0.475\linewidth,valign=m]{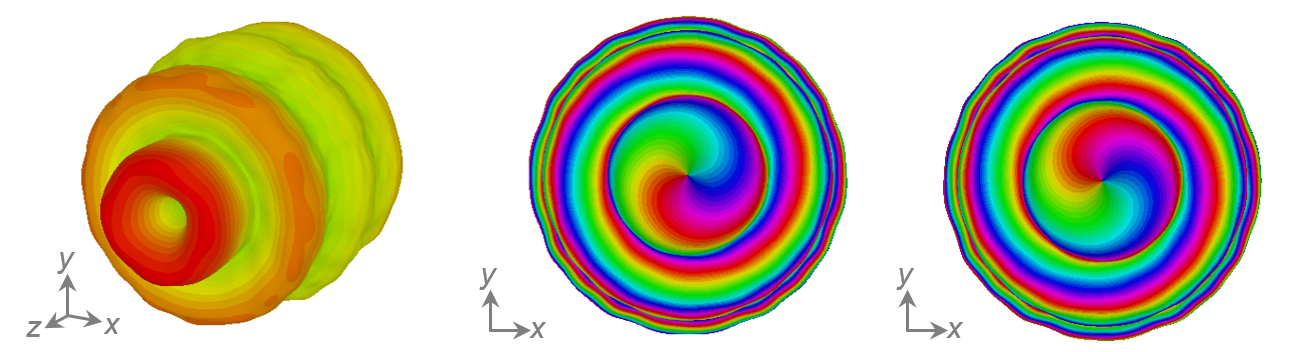}
		%		\vphantom{
		%			\includegraphics[width=0.3\linewidth,valign=m]{Fig004B.pdf}
		%		}
		\label{fig6B}
	}
	\\
	\subfloat[$ \ell =\pm2 $]{
		\includegraphics[width=0.475\linewidth,valign=m]{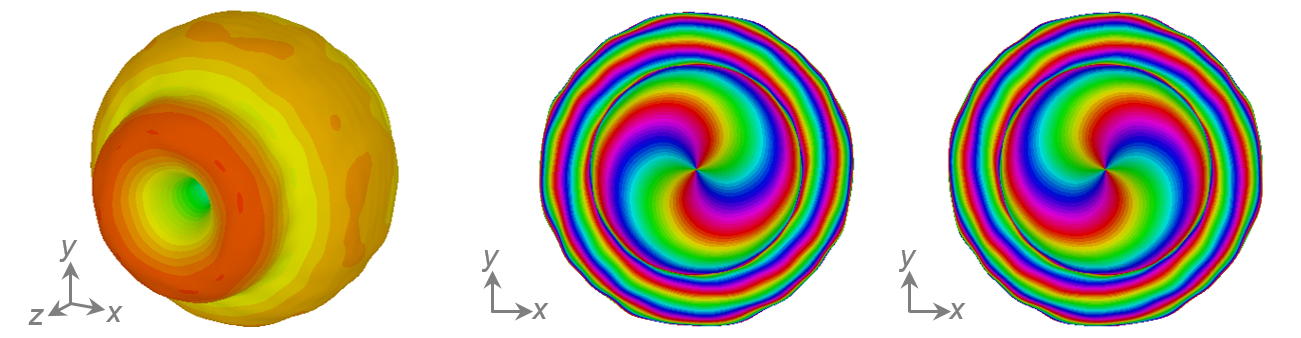}
		\label{fig6C}
	}
	\hfil
	\subfloat[$ \ell =\pm3 $]{
		\includegraphics[width=0.475\linewidth,valign=m]{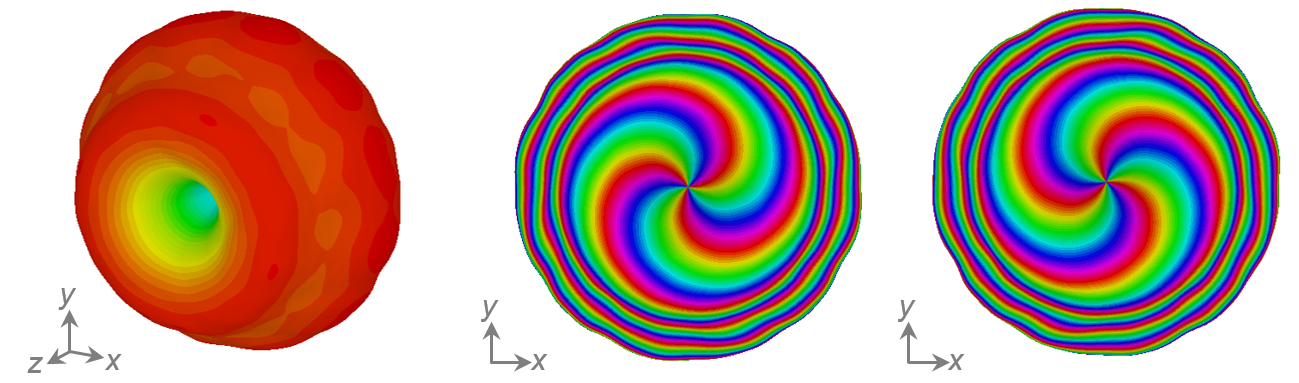}
		\label{fig6D}
	}
	\\
	\subfloat[$ \ell =\pm4 $]{
		\includegraphics[width=0.475\linewidth,valign=m]{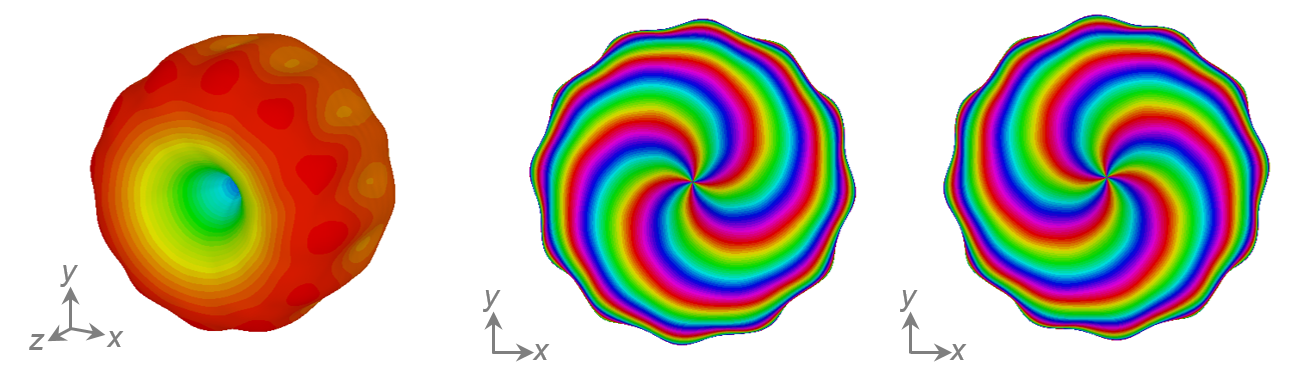}
		\label{fig6E}
	}
	\hfil
	\subfloat[$ \ell =\pm5 $]{
		\includegraphics[width=0.475\linewidth,valign=m]{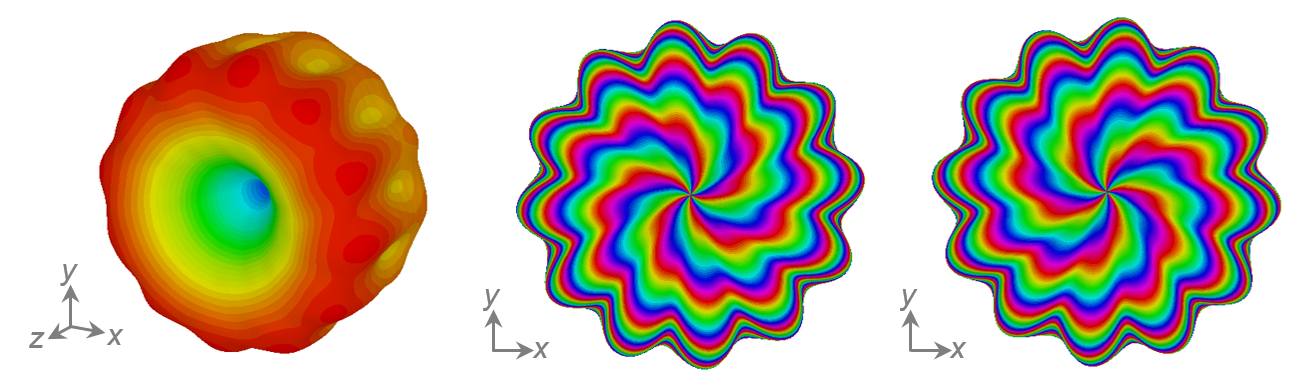}
		\label{fig6F}
	}
	%	\\
	%	\subfloat[$ \ell =\pm6 $]{
	%		\includegraphics[width=0.45\linewidth,valign=m]{Fig006G.PNG}
	%		\label{fig6G}
	%	}
	%	\hfil
	%	\subfloat[$ \ell =\pm7 $]{
	%		\includegraphics[width=0.45\linewidth,valign=m]{Fig006H.PNG}
	%		\label{fig6H}
	%	}
	\caption{ Predicted normalized far--field directivity patterns of OAM modes (a) $\ell = 0$, (b) $\ell = \pm1$, and (c) $\ell = \pm2$, (d) $\ell = \pm3$, (e) $\ell = \pm4$ and (f) $\ell = \pm5$.
		%		, (g) $\ell = \pm6$, and (h) $\ell = \pm7$. 
		%
		The phase/magnitude colour map scales and axis shown in (a) are common for all the OAM modes.}
	\label{fig6}
\end{figure*}
\begin{figure*}[!t]
	\centering
	\subfloat[]{
		\includegraphics[width=0.475\linewidth,valign=m]{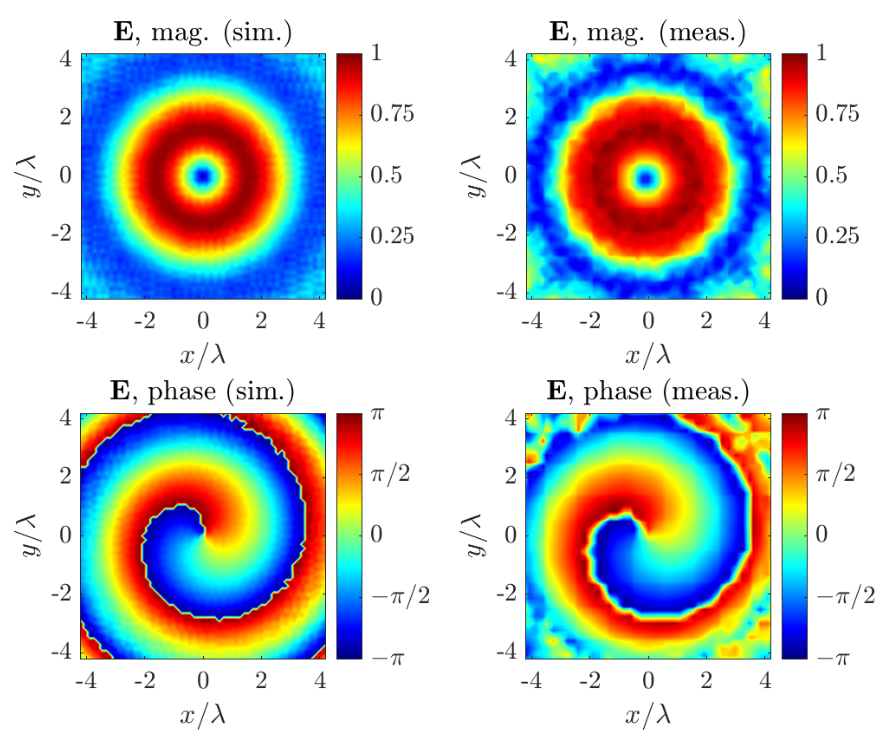}
		\label{fig7A}
	}
	\hfil
	\subfloat[]{
		\includegraphics[width=0.475\linewidth,valign=m]{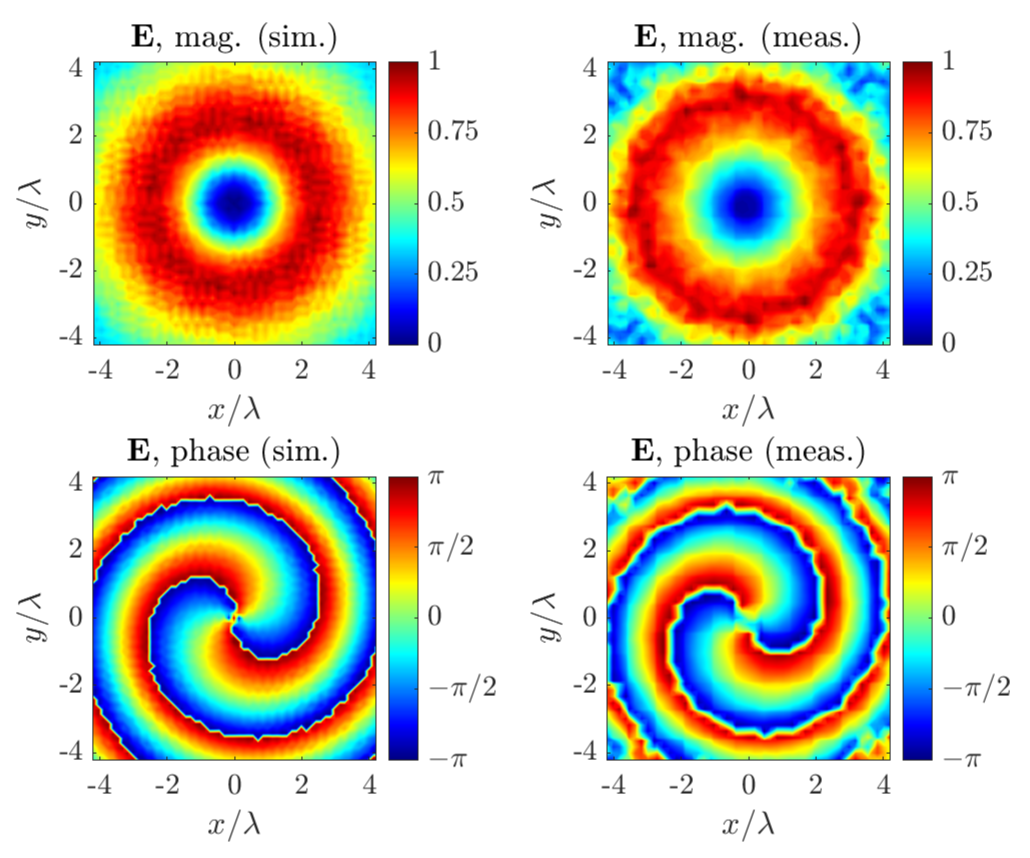}
		\label{fig7B}
	}
	\caption{ Comparison between measured and simulated near-field electric field patterns of the OAM modes (a) $\ell = +1$ and (b) $\ell = +2$: a good agreement is observed.}
	\label{fig7}
\end{figure*}

Fig. \ref{fig6} depicts the simulated results.
Excitation of mode $\ell = 0 $ (Fig. \ref{fig6A}) reveals a directional radiation pattern with constant phase value around the $+z$-direction. 
When modes $\ell = \pm 1$ and $\ell = \pm2$ are excited (Fig. \ref{fig6B} and \ref{fig6C}, respectively), vortices are revealed along the $+z$-direction, while single and double cycles of phase spiral ($ 0 $ deg.--$ 360 $ deg.) are observed in the radiation pattern phase. The phase spiral is clockwise for $\ell = +1, +2$ while it is counter-clockwise for $\ell = -1 , -2$. 
Note that the phase response for a given $\ell$-th modes is responsible for spatial orthogonality, hence simultaneous data transmission using OAM modes is possible, as verified by literature \cite{Chen2018,Wang2019,4,5,6,7,8,10,12}. 
The far-field response against the remaining $ 6 $ out of $ 15 $ mode excitations is shown in Fig. \ref{fig6D} to \ref{fig6F}. Other than the excitation of mode $\ell = 0$ (Fig. \ref{fig6A}), all other mode excitations create vortices along $+z$-direction. 
The number of $ 0 $ deg.--$ 360 $ deg. clockwise and anti-clockwise phase spirals increases as we move higher in the mode number (i.e. $\ell = \pm 1 ... \pm 5$), and this trend breaks for the highest OAM modes i.e. $\ell = \pm 5$. 
%
%It can be observed that the number of phase spirals generated for mode $\ell = + 7$ excitation is identical to that of $\ell = -5$. The same is true for the mode excitation $\ell = -7$ and $\ell = +5$. 
%
%Even with this resemblance of the radiation patterns phase, the radiation pattern magnitude for mode excitation $\ell = \pm5$ and $\ell = \mp7$ is different. 
%
Another interesting feature to observe is that the radiation pattern magnitude for mode excitation $\ell = \pm3, \pm4$ and $\pm5$ reveal quasi-broadcast transmission along $xy$-plane in addition to the OAM modes in $+z$-direction, which provides an additional flexibility in terms of application scenarios to the presented circular array. 
%In a similar fashion, radiation pattern magnitude for mode excitation $\ell = \pm6$ and $\pm7$ show quasi-multicast transmission. 

\begin{figure*}[!t]
	\centering
	\includegraphics[width=0.5\linewidth]{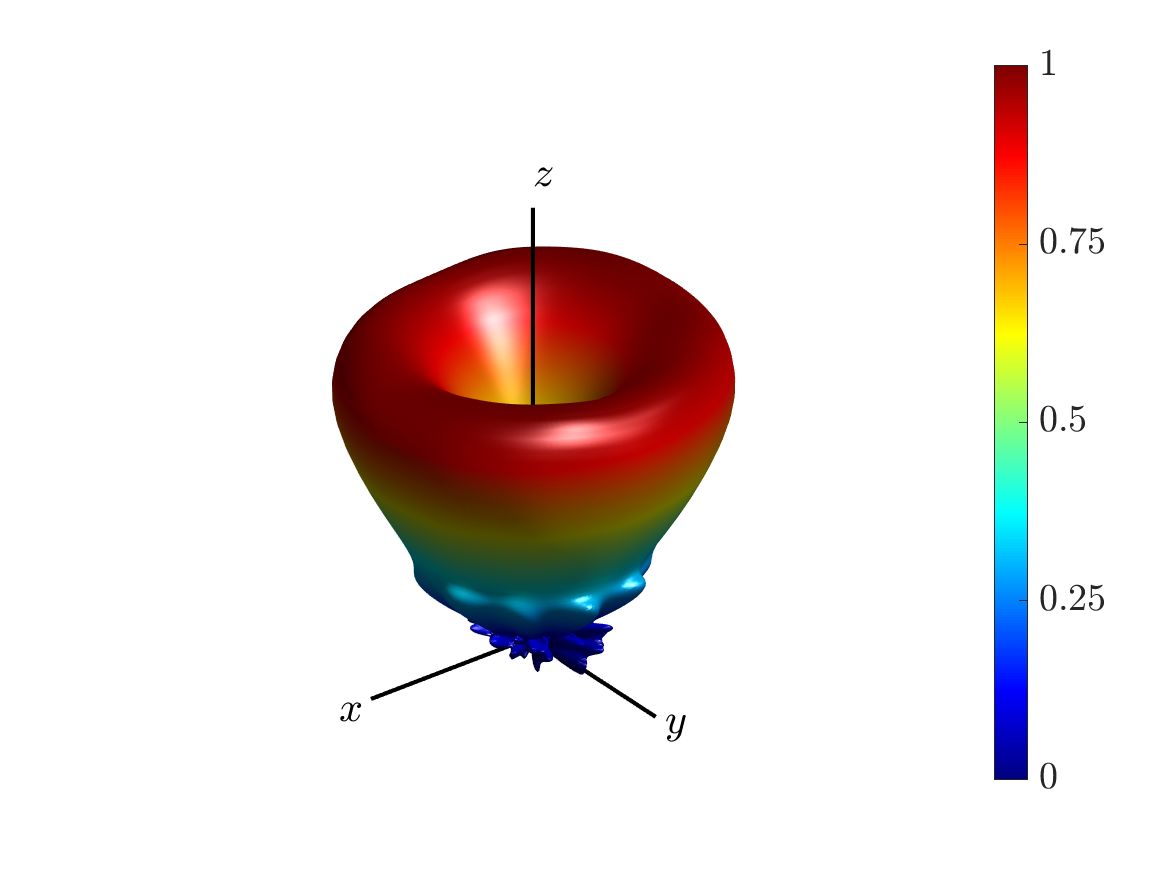}
	\caption{Normalized far--field radiation pattern (3D representation) of the circular antenna array, generated by the near--field measurements, estimated through the near-- to far--field transformation (based on the \textit{asymptotic evaluation} method), for the first OAM mode ($ \ell=1 $).}
	\label{fig9}
\end{figure*}

Two of the OAM modes ($ \ell=1,2 $) are verified by measurements in planar near-field anechoic chamber. The circular array is placed in near-field of a $ 28 $ GHz scanning probe such that the probe is facing the $+z$-direction of the circular array placed in $xy$-plane. Electric field is recorded, normalized to the maximum field value, and is compared to the predictions from CST Microwave Studio simulations. The results are presented in Fig. \ref{fig7}. For healthy comparison, the phase spiral for simulation is rotated in a post-processing step to match with the phase spiral recorded from measured data. Plots against $\ell = +1$ excitation in Fig. \ref{fig7A} show that the phase spiral is matched well with the simulated predictions while the magnitude is flattened around the vortex in $+z$-direction. 
%
%This means that the measured electric field pattern is not as high as it was predicted in simulations, while OAM mode generation is verified. 
%
In the same way, phase spiral matches well with the simulated results for $\ell = +2$ excitation. 
All the above measured results verify the OAM function of the antenna array in the vertical direction.
Finally, the far-field radiation pattern is estimated from the near field measurement results, through the near to far--field transformation, based on the \textit{asymptotic evaluation} method, described in \cite{balanis2016antenna}, and the results (3D representation of the normalized magnitude in spherical coordinates) is depicted in Fig. \ref{fig9}.

\section*{Methods}
\subsection*{Simulation setup}
The antenna was simulated in terms of reflection coefficient, cross coupling and radiation pattern through full electromagnetic analysis by using the commercial  numerical solver CST Microwave Studio suite. Specifically, the frequency domain solver   (finite element method (FEM)) was used. Discrete ports at $ 50 $ ohm were applied as excitation. In order to estimated the radiation pattern for each mode, for the uni-- and multi--cast and for the OAM transmission, the schematic representation of the antenna and the \textit{AC, combine result simulation task} tool was used.

\subsection*{Measurement setup}
\begin{figure*}[!t]
	\centering
	\subfloat[]{
		\includegraphics[height=0.3\linewidth,valign=m]{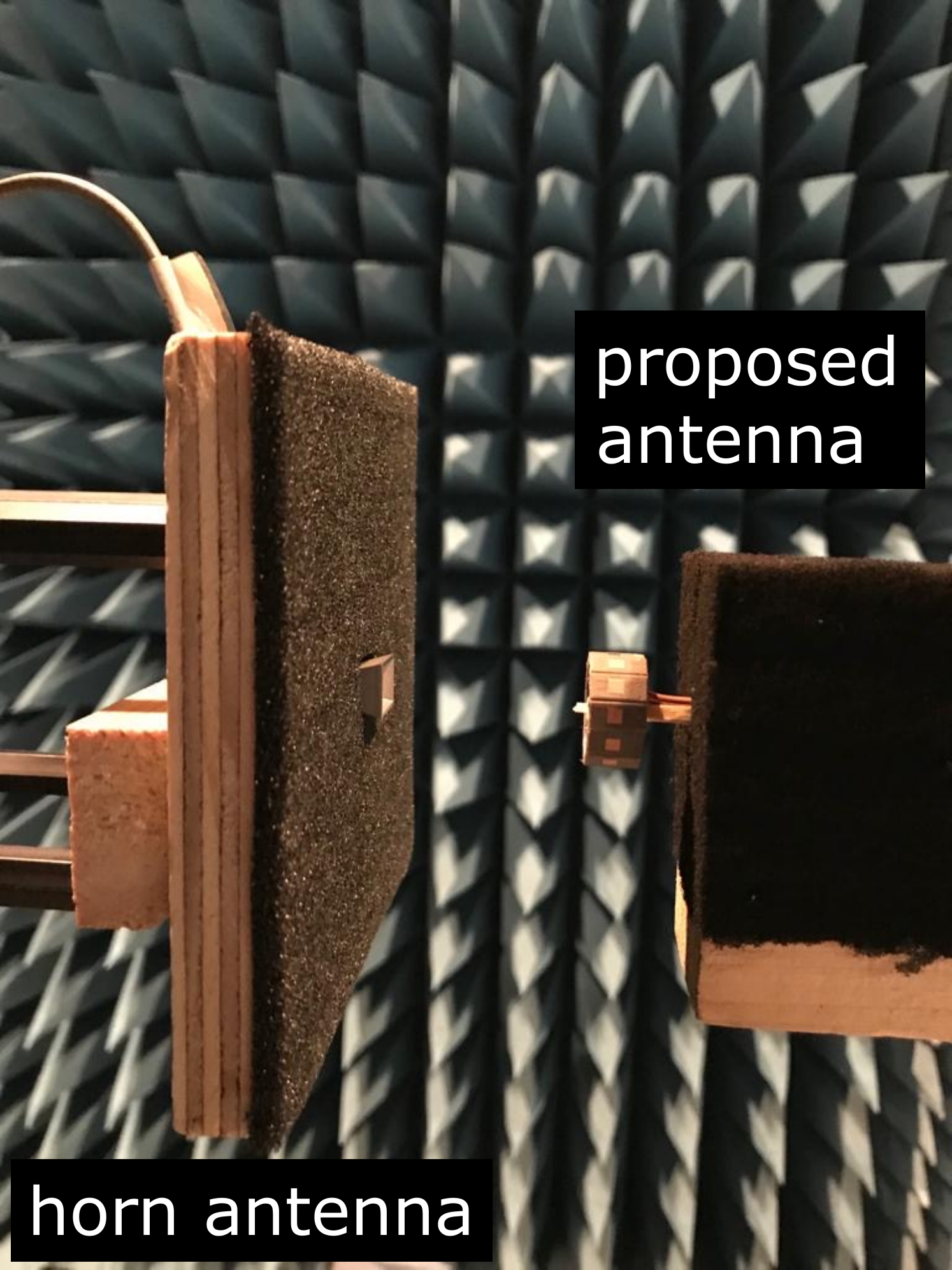}
		\label{fig10A}
	}
	\hfil
	\subfloat[]{
		\includegraphics[height=0.3\linewidth,valign=m]{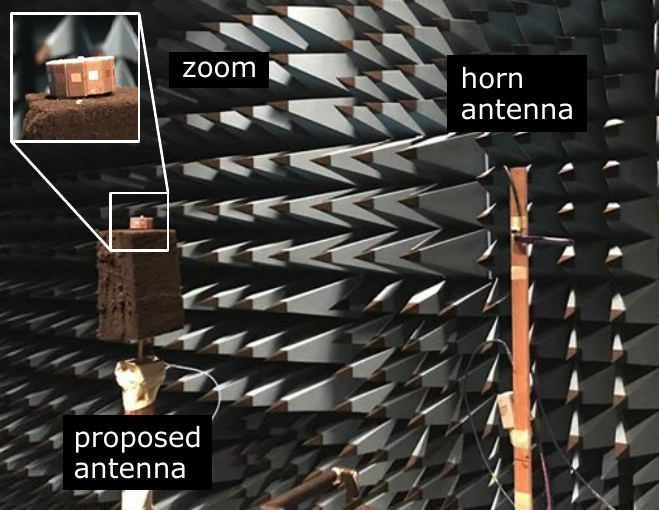}
		\label{fig10B}
	}
	\caption{Measurement setup  for the near-- (a) and far--field (b) measurements.}
	\label{fig10}
\end{figure*}

The measurement setup for the near-- and far--field is depicted in Fig. \ref{fig10}, respectively. 
For the near--field measurements  the proposed antenna was placed at vertical distance of $ 4\lambda $ from a receiver horn antenna. 
The latter scanned a planar region with dimensions $ 4\lambda \times 4\lambda $ and captured the transmitted by the proposed antenna electric field. The region was divided into $ 61\times{}61 $ measurement points.
For the cross--polarized component measurement the horn antenna was rotated by $ 90 $ deg.
For the far-field measurement, the receiver horn antenna is placed at $ 100\lambda$ distance. Both antennas lie in the horizontal plane. Now, the proposed antenna rotates $ 360 $ degrees around its axis and the horn antenna captures the transmitted signal.

\section*{Conclusion}
This work presents a circular patch antenna array capable of generating uni- and multi-cast transmission along azimuth plane and, simultaneously, a high number of OAM modes in elevation direction. 
The mode-based analysis is thoroughly investigated through fundamental principles and it is shown that mode-mixing of circular antenna array can result in various uni-cast and multi-cast transmissions. 
OAM mode excitations can result in several OAM modes transmission, where most of the latter show high level of spatial orthogonality. 
Antenna was fabricated via a conventional  milling machine and tested in terms of reflection coefficient, near- and far-field radiation patter.
Simulated results through full electromagnetic analysis and measured results agree very well in all cases.
The array is perfect candidate for high spectral efficiency data transmission for 5G and beyond wireless applications. 

\section*{Acknowledgment}

This research was supported in part by the UK Engineering and Physical Sciences Research Council (EPSRC) under the UKRI grants EP/P000673/1 and EP/NO20391/1.
Authors would like to thank K. Rainey and O. Malyuskin for assisting in the hardware development and measurements.

\section*{Author Contributions}
S.D.A. and V.F. conceived the idea. S.D.A. designed and simulated the proposed antenna, performed the measurements, interpreted results and edit the paper. M.A.B.B wrote the paper. S.D.A. and V.F. supervised the research and contributed to the general concept and interpretation of the results. All authors reviewed the manuscript.

\balance

%\bibliography{referenses001}
	
%	
	\bibliographystyle{IEEEtran}
	\bibliography{referenses001.bib}

\end{document}